\documentclass[11pt,a4paper]{article}
\pdfoutput=1

\usepackage{amsfonts}
\usepackage{amssymb}
\usepackage{mathrsfs}
\usepackage{amsmath,amssymb}
\usepackage{subeqn}
\usepackage{bbm}
\usepackage[bookmarks=true]{hyperref}
\usepackage[nosort]{cite}
\usepackage[bulletsep]{collref}

\usepackage[british]{babel} %dates in british format

\usepackage{color}

\usepackage{epsfig,psfrag,graphicx}

\ifx\hypersetup\sadfkjashdfkxja\else
\hypersetup{pdftitle={Shaping up BPS States with Matrix Model Saddle Points}}
\hypersetup{pdfauthor={Diego Correa, Martin Wolf}}
\hypersetup{pdfsubject={Mathematical Physics}}
\hypersetup{pdfkeywords={ }}
\hypersetup{plainpages=false}
\hypersetup{bookmarksnumbered=true}
\hypersetup{pdfstartview=FitH}
\hypersetup{pdfpagemode=None}
\hypersetup{colorlinks=false}
\hypersetup{citebordercolor={.5 1 .5}}
\hypersetup{urlbordercolor={.5 1 1}}
\hypersetup{linkbordercolor={1 .7 .7}}
\fi

\makeatletter\renewcommand{\section}{\@startsection
{section}{1}{\z@}{-2.5ex plus -1ex minus
    -.2ex}{2.3ex plus .2ex}{\centering\large\bf\mathversion{bold}}}

\makeatletter\renewcommand{\subsection}{\@startsection{subsection}{2}{\z@}{
-3.25ex
plus -1ex minus
   -.2ex}{1.5ex plus .2ex}{\centering\bf\mathversion{bold}}}

\makeatletter\renewcommand{\subsubsection}{\@startsection{subsubsection}{3}{
-2.45ex}{-3.25ex
plus -1ex minus -.2ex}{1.5ex plus .2ex}{\centering\bf\mathversion{bold}}}

\makeatletter\renewcommand{\paragraph}{\@startsection{paragraph}{4}{\z@}%
                                    {0.8ex \@plus1ex \@minus.2ex}%
                                    {-.5em}%

{\normalfont\normalsize\bfseries\mathversion{bold}}}

\renewcommand{\thesection}{\arabic{section}.}
\renewcommand{\thesubsection}{\arabic{section}.\arabic{subsection}.}

\numberwithin{paragraph}{section}
\renewcommand\theparagraph {\S\thesection\@arabic\c@paragraph.\kern-8pt}

\numberwithin{equation}{section}
\renewcommand{\theequation}{\thesection\arabic{equation}}

\renewcommand*\l@section{\@dottedtocline{1}{0em}{1.5em}}
\renewcommand*\l@subsubsection{\@dottedtocline{4}{3.8em}{3.2em}}

\renewcommand\tableofcontents{%
    \section*{\large\contentsname
        \@mkboth{%
          \MakeUppercase\contentsname}{\MakeUppercase\contentsname}}%
       {\baselineskip=15pt plus 2pt minus 1pt
    \@starttoc{toc}}%
}

\renewenvironment{thebibliography}[1]
     {\section*{\large\centering{\refname}
        \@mkboth{\MakeUppercase\refname}{\MakeUppercase\refname}}%
     \list{\@biblabel{\@arabic\c@enumiv}}%
           {\settowidth\labelwidth{\@biblabel{#1}}%
            \leftmargin\labelwidth
            \advance\leftmargin\labelsep
            \@openbib@code
            \usecounter{enumiv}%
            \let\p@enumiv\@empty
            \renewcommand\theenumiv{\@arabic\c@enumiv}}%
      \sloppy
      \clubpenalty4000
      \@clubpenalty \clubpenalty
      \widowpenalty4000%
      \sfcode`\.\@m
 \catcode`\^^M=10%
}

\DeclareFontFamily{U}{rsf}{}
\DeclareFontShape{U}{rsf}{m}{n}{
  <5> <6> rsfs5 <7> <8> <9> rsfs7 <10-> rsfs10}{}
\DeclareMathAlphabet\Scr{U}{rsf}{m}{n}

\newcommand{\dd}{\mathrm{d}}
\newcommand{\di}{\mathrm{i}}
\newcommand{\de}{\mathrm{e}}

\newcommand{\ewith}{\quad\mbox{with}\quad}
\newcommand{\eand}{\quad\mbox{and}\quad}
\newcommand{\efor}{\quad\mbox{for}\quad}

\newcommand{\IR}{\mathbbm{R}}
\newcommand{\IZ}{\mathbbm{Z}}

\newcommand{\IN}{\mathbbm{N}}

\newcommand{\cH}{\mathscr{H}}

\newcommand{\cZ}{\mathscr{Z}}

\newcommand{\CN}{\mathcal{N}}

\newcommand{\sSU}{\mathsf{SU}}

\newcommand{\be}{{\beta}}

\newcommand{\cor}{\color{red}}

\def\be{\begin{equation}}
\def\ee{\end{equation}}
\def\ba{\begin{eqnarray}}
\def\ea{\end{eqnarray}}
\def\bc{\begin{center}}
\def\ec{\end{center}}

\begin{document}

\begin{titlepage}

\setcounter{page}{0}
\renewcommand{\thefootnote}{\fnsymbol{footnote}}

\begin{flushright}
DAMTP 2010--56\\[.5cm]
\end{flushright}

\vspace*{1cm}

\begin{center}

{\LARGE\textbf{Shaping up BPS States\\ with Matrix Model Saddle Points}\par}

\vspace*{1cm}

{\large
 Diego H.~Correa and Martin Wolf\footnote{Also at the Wolfson College,
 Barton Road, Cambridge CB3 9BB, United Kingdom.} \footnote{{\it E-mail
addresses:\/}
\href{mailto:d.correa@damtp.cam.ac.uk}{\ttfamily d.correa@damtp.cam.ac.uk},
\href{mailto:m.wolf@damtp.cam.ac.uk}{\ttfamily m.wolf@damtp.cam.ac.uk}
}}

\vspace*{1cm}

{\it Department of Applied Mathematics and Theoretical Physics\\
University of Cambridge\\
Wilberforce Road, Cambridge CB3 0WA, United Kingdom}

\vspace*{1cm}

{\bf Abstract}
\end{center}
\vspace*{-.3cm}

\begin{quote}
We provide analytical results for the probability distribution of a family of
wavefunctions of a quantum mechanics model of commuting matrices in the
large-$N$ limit. These wavefunctions describe the strong coupling limit of 1/8
BPS states of $\CN=4$ supersymmetric Yang--Mills theory. In the large-$N$
limit, they should be dual to classical solutions of type IIB supergravity that
asymptotically approach AdS$_5\times S^5$. Each probability distribution can be
described as the partition function of a matrix model (different wavefunctions
correspond to different matrix model potentials) which we study by means of a
saddle point approximation. These saddle point solutions are  given in terms of
(five-dimensional) hypersurfaces supporting density distributions of
eigenvalues.

\vfill
\noindent 29th July 2010

\end{quote}

\setcounter{footnote}{0}\renewcommand{\thefootnote}{\arabic{thefootnote}}

\end{titlepage}

\tableofcontents

\bigskip
\bigskip
\hrule
\bigskip
\bigskip

\section{Introduction}

Since the discovery of the anti-de Sitter space/conformal field theory
(AdS/CFT)
correspondence \cite{Maldacena:1997re,Gubser:1998bc,Witten:1998qj}, significant
progress has been made in our understanding of strongly coupled gauge theory
phenomena. So far, the example that has been studied most extensively is the
correspondence between $\sSU(N)$ maximal $\CN=4$ supersymmetric Yang--Mills
(SYM) theory in four dimensions and  type IIB superstring theory on the
ten-dimensional AdS$_5\times S^5$ background. The correspondence in its
strongest form claims full dynamical agreement between both theories at the
quantum level. This is certainly hard to verify and one therefore seeks limits
in which  it is possible to perform tests explicitly. One very interesting
limit
is the planar or large-$N$ limit of the gauge theory. In this limit, $\CN=4$
SYM
theory is believed to be equivalent to free (i.e.~genus-zero) string theory on
AdS$_5\times S^5$. If one in addition assumes strong 't Hooft coupling (or
equivalently, large curvature radius of AdS), then the gauge theory is believed
to be described by classical supergravity.

One may adopt the point of view of taking the AdS/CFT correspondence as an
approach towards defining a theory of quantum gravity. This then naturally
leads to the question of how geometrical information emerges in the strong
coupling limit of the gauge theory. Notice that gravity is not  apparent in the
Lagrangian description of quantum field theory and in this sense it can be
thought of as `emergent'. Moreover, the gravity side of the correspondence is
higher dimensional, therefore one should also be able to understand how
excitations are localised in the dual extra dimensions by field theoretical
means.

To shed light on the problem of emergent geometry, one needs to study strongly
coupled gauge theory, which on general grounds, is extremely hard. To tackle
this problem, Berenstein \cite{Berenstein:2005aa} proposed a truncation of
$\CN=4$ SYM to a quantum mechanical problem of commuting matrices by
compactifying the theory on a three-sphere (i.e.~he considered $\CN=4$ SYM
theory on $\IR\times S^3$). This compactification provides a natural infra-red
regulator. Moreover, upon expanding all the fields of $\CN=4$ SYM theory in
spherical harmonics on the three-sphere, the action is truncated  to obtain a
quantum mechanics Hamiltonian of six Hermitian matrices. This truncation of the
degrees of freedom to commuting matrices is a good approximation in the strong
coupling limit.\footnote{The notion of emergent geometry in the strong coupling
limit of  more general matrix models was studied in \cite{Berenstein:2008eg}.}
The eigenstates of the model with six commuting matrices are
conjectured to describe 1/8 BPS states of $\CN=4$ SYM theory at strong
coupling.
In the large-$N$ limit, these are dual to classical solutions of type IIB
supergravity that asymptotically approach AdS$_5\times S^5$.  Similar models
have been proposed to describe BPS states in orbifolds of $\CN=4$ SYM
theory \cite{Berenstein:2005ek,Berenstein:2006yy} and certain $\CN=1$
superconformal field theories \cite{Berenstein:2007wi,Berenstein:2007kq}.

A Gaussian wavefunction of the eigenvalues of the commuting matrices was shown
to be the exact ground state of the quantum mechanical Hamiltonian
\cite{Berenstein:2005aa}. The product of this ground state by a holomorphic
function of the matrices' eigenvalues is an eigenfunction of the Hamiltonian,
as
well, at least to a good approximation in the large-$N$ limit
\cite{Berenstein:2007wi,Berenstein:2007kq,Berenstein:2007wz}.
Now, having this family of quantum mechanical eigenstates, it would be
desirable
to characterise their typical or most likely distribution of eigenvalues. The
probability distribution of each of the wavefunctions can be seen as the
partition function of a given matrix model. For the ground state, the resulting
partition function would be that of a model with a quadratic potential and a
generalised Vandermonde repulsion. For the `excited' wavefunctions, other terms
are added to the potential of the model. In the large-$N$ limit, all these
partition functions will be dominated by their saddle points. Although only for
the ground state the saddle point equations have been solved exactly so far
\cite{Berenstein:2005aa,Berenstein:2005jq}, the result is very compelling: The
saddle point configuration is a uniform distribution of eigenvalues supported
on a five-sphere embedded into $\IR^6$. Just to remind the reader, the
ground state wavefunction must be identified with the dual of the AdS$_5\times
S^5$ background. Subsequent studies revealed that it is possible to confer
an explicit geometrical interpretation to the five-sphere of eigenvalues
\cite{Berenstein:2005jq}. In order to see if one can push this identification
further, it would be necessary to compute the partition functions associated
with other eigenstates of the quantum mechanical Hamiltonian. The simplest
cases
one could start from to look at are  1/2 BPS eigenstates. For these states the
AdS/CFT dictionary has been studied thoroughly (see
e.g.~\cite{Corley:2001zk,Berenstein:2004kk,Takayama:2005yq,Donos:2005vm,
Koch:2009gq}) and it is known how to relate them to a family of 1/2 BPS
supergravity solutions that asymptotically approach AdS$_5\times S^5$, found by
Lin, Lunin \& Maldacena (LLM) \cite{Lin:2004nb}\footnote{An alternative
approach
to recast geometrical information of LLM solutions was used in
\cite{Vazquez:2006id,Chen:2007gh,Koch:2008ah,deMelloKoch:2009zm,Lin:2010sb}.}.
Unfortunately, solving the corresponding partition functions in the quantum
mechanics model with six commuting matrices is a very difficult problem, even
in
the saddle point approximation. The only reported results in this direction are
in a series of articles that study some of these wavefunctions numerically for
finite $N$ \cite{Berenstein:2007wz,Berenstein:2008jn,Berenstein:2010dg}.

In this paper, we will extend the analytical result that is known for the
ground
state wavefunction to excited wavefunctions. Specifically, we will develop a
perturbative method that allows for an analytical treatment of the saddle point
equations of a family of wavefunctions. We will present analytical results for
monomial and logarithmic potentials in the corresponding matrix model
Hamiltonian. Wavefunctions for degree $p\in\IN$ monomial potentials are
expected to be in correspondence with LLM geometries obtained from a simply
connected droplet possessing a non-vanishing $p^{\rm th}$ harmonic moment
\cite{Vazquez:2006id}. Wavefunctions for logarithmic potentials should
correspond to annular LLM geometries where the inner radius is determined by
the
strength of the logarithmic potential \cite{Berenstein:2005aa}.  Our
results should form a starting point for extracting geometric information
directly from the gauge theory. We will comment on this issue in the
conclusions.

This paper is organised a follows. We will first give a brief review of the
background material. In Section \ref{sec:Analytical}, we then provide our
perturbative approach and present analytical solutions for monomial and
logarithmic potentials. We shall also compare our results against the numerics.
In Section \ref{sec:Conclusions}, we will conclude and give an outlook of open
problems. Finally, several appendices collect useful definitions and details of
our derivations.

\section{Quantum mechanics of commuting matrices}\label{sec:MQM}

The system we are going to be dealing with is a particular matrix quantum
mechanics model of six commuting Hermitian matrices. Let $\vec
X=(X^1,\ldots,X^6)$ be the six Hermitian matrices, they are therefore subject
to
the constraint
\begin{equation}\label{eq:Commuting}
 \vec X\wedge\vec X\ =\ 0~.
\end{equation}
The Hamiltonian we are interested in is
\begin{equation}
 H^{\rm cl}
 \ =\ \tfrac12\,\mbox{tr}\big(\vec\Pi\cdot\vec\Pi+\vec X\cdot\vec X\big)~,
\end{equation}
where $\vec\Pi$ is conjugate to $\vec X$ and `cl' refers to classical. The
system has a gauge invariance, where one acts by conjugation: $\vec X\mapsto
g^{-1}\vec Xg$ for $g\in\sSU(N)$. Because of the constraint
\eqref{eq:Commuting}, one can use this $\sSU(N)$ action
to diagonalise all six matrices $\vec X$ simultaneously. Let us denote the
eigenvalues of $\vec X$ by $\vec x_i=(x^1_i,\ldots,x^6_i)\in\IR^6$
for $i,j,\ldots=1,\ldots,N$. Having diagonalised all the matrices, we have
fixed a gauge. However, there are still residual gauge transformations which
permute the eigenvalues $\vec x_i$. Therefore, the corresponding wavefunctions
will eventually be symmetric under the exchange $\vec x_i\leftrightarrow\vec
x_j$ for all $i$ and $j$.

As shown by Berenstein \cite{Berenstein:2005aa}, this system can be obtained
as a truncation of $\CN=4$ SYM theory  on $\IR \times S^3$, to the $s$-wave
modes of its six scalar fields, to describe gauge invariant $1/8$ BPS states in
the strong coupling limit.

Having reduced the dynamics of these six $N\times N$ matrices to the dynamics
of their eigenvalues, the system can thus be interpreted as set of $N$ bosons
on a space with six dimensions. If we treat the system classically, we can use
a
diagonal ansatz to find solutions of the dynamical system. Under these
assumptions, we find $N$ free harmonic oscillators in six dimensions, which
should be treated as $N$ identical particles (bosons) on a six-dimensional
harmonic oscillator.

Quantum mechanically, we cannot do that immediately. This is due to a certain
measure factor that arises from the volume of the gauge orbit, and which
affects
the dynamics of the system. This measure factor was computed in
\cite{Berenstein:2005aa} and it is given by
\begin{equation}
\mu^2 \ :=\ \prod_{1\leq i<j\leq N} |\vec x_i-\vec x_j|^2~.
\end{equation}
The resulting quantum Hamiltonian is therefore
\begin{equation}
 H\ =\ \sum_{i=1}^N\left(-\frac 1{2\mu^2}\, \vec\nabla_i \cdot \mu^2\vec
\nabla_i + \frac 12\, |\vec x_i|^2\right).
\label{eq:HQM}
\end{equation}

\subsection{Wavefunctions}

The main object of study in this paper will be some wavefunctions of the
Hamiltonian \eqref{eq:HQM}. The presence of the Vandermonde measure factor
makes
the corresponding Schr\"odinger problem very difficult. Notice nonetheless that
the rather simple wavefunction
\begin{equation}
 \psi_0(\vec x_1,\ldots,\vec x_N)\ =\ \exp\left(-\frac{1}{2}\sum_{i=1}^N |\vec
x_i|^2 \right)
\end{equation}
is an exact wavefunction of $H$. In fact, it is the ground state wavefunction,
\begin{equation}
 H\psi_0\ =\ E_0\psi_0~,\ewith E_0\ =\ \frac{N}{2}(N-1)+3N~.
\end{equation}

The measure factor $\mu$, which will appear in the probability density
distribution, can be absorbed into the wavefunction $\psi$ by a similarity
transformation
\begin{equation}
 \psi\ \mapsto\ \hat \psi\ :=\ \mu \psi~.
\end{equation}
In the following, we shall concern ourselves with the re-scaled wavefunctions
$\hat\psi$ only.

If we square $\hat \psi$, we get a probability  density distribution on the
phase space of the $N$ particles. For the ground state $\hat\psi_0$, this is
given by
\begin{equation}
 |\hat \psi_0|^2\ =\  \mu^2 \exp\left(-\sum_{i=1}^N |\vec x_i|^2\right)
 \ =\ \exp \left(-\sum_{i=1}^N |\vec x_i|^2+ \sum_{1\leq i<j\leq N}
 \log|\vec x_i-\vec x_j|^2\right).
\end{equation}
If we set  $\cH_0:=-\log|\hat \psi_0|^2$, then
\begin{equation}
 \cZ_0\ =\ \int\left(\prod_{i=1}^N\dd^6 x_i\right)\exp(-\beta\cH_0)~,
 \ewith \beta\ \equiv\ 1
\end{equation}
can be interpreted as the partition function of a gas of particles in a
confining external quadratic potential, $\sum_i|\vec x_i|^2$, together with a
logarithmic repulsion term, $-\sum_{i<j}\log |\vec x_i-\vec x_j|^2$, between
the
particles in six dimensions.\footnote{This logarithmic repulsion generalises
the Vandermonde repulsion of eigenvalues in matrix models of
\cite{Brezin:1977sv}.}

Less is known about the exact excited wavefunctions for the Hamiltonian $H$.
However, in \cite{Berenstein:2005aa,Berenstein:2007wz} it was  shown that for
$1/2$ BPS states, the wavefunction
\begin{equation}
 \hat\psi_f(\vec x_1,\ldots,\vec x_N)\ :=\ \exp\left(\sum_{i=1}^N
f(z_i)\right)\hat\psi_0(\vec x_1,\ldots,\vec x_N)
\end{equation}
is an approximate eigenfunction of $H$ in the thermodynamic limit $N\to
\infty$,
provided $f=f(z_i)$ is a holomorphic function of $z_i:=x_i^5+\di x_i^6$. The
partition function in this case is then given by
\begin{subequations}
 \begin{equation}\label{eq:partitionfunction}
   \cZ_f\ =\ \int\left(\prod_{i=1}^N \dd^6 x_i\right) \exp(-\beta\cH_f)~,
  \ewith\beta\ \equiv\ 1~,
 \end{equation}
where
\begin{equation}\label{eq:Hdiscrete}
 \cH_f\ :=\ -\log |\hat\psi_f|^2\ =\
       \sum_{i=1}^N |\vec x_i|^2 -2\sum_{i=1}^N\mathfrak{Re}f(z_i)-
\sum_{1\leq i<j\leq N}\log|\vec x_i-\vec x_j|^2~.
\end{equation}
\end{subequations}
Here, `$\mathfrak{Re}$' denotes the real part.

In this work, we shall be interested in the large-$N$ limit of the  `matrix
model' partition function \eqref{eq:partitionfunction}. In this limit, the
bosons will form some type of distribution density $\rho$ on the phase space of
a single particle (density of eigenvalues). The goal for us is then to
determine
the shape of the density $\rho$  using
a saddle point approximation.

\subsection{Large-$N$ limit}

Next we wish to perform the thermodynamic limit of the Hamiltonian
\eqref{eq:Hdiscrete}. When taking the limit $N\to \infty$, we may trade the
sums in $\cH_f$ for integrals at the expense of introducing a density
$\rho=\rho(\vec x)$ (which is constrained to be non-negative),
\begin{subequations}
\begin{equation}
 \sum_{i=1}^N\ \rightarrow\  \int\!\dd^6x\, \rho(\vec x)\eand
 \sum_{1\leq i<j\leq N} \ \rightarrow\ \frac12\!\int\!\dd^6x\,\dd^6x'
\rho(\vec x) \rho(\vec x')~,
\end{equation}
and which is subject to the normalisation
\begin{equation}\label{eq:Normalisation}
 \int\!\dd^6x\,\rho(\vec x)\ =\ N~.
\end{equation}
\end{subequations}
Therefore, \eqref{eq:Hdiscrete} becomes
\begin{equation}\label{eq:HContinuous}
 \cH_f\ =\ \int\!\dd^6x\, \rho(\vec x)|\vec x|^2  -2\int\!\dd^6x\,
 \rho(\vec x)\,\mathfrak{Re} f(z)-\frac{1}{2}\!\int\!\!\int\!\dd^6x\,\dd^6x'
 \rho(\vec x) \rho(\vec x') \log |\vec x-\vec x'|^2~.
\end{equation}
Notice that the constraint \eqref{eq:Normalisation} might be added to
\eqref{eq:HContinuous} by using a Lagrange multiplier $\Lambda$.

In the large-$N$ limit, the partition function (\ref{eq:partitionfunction})
will be dominated by its saddle point. Then, the most likely density $\rho$ can
then be obtained by extremising the Hamiltonian $\cH_f$. Specifically, the
variation of $\cH_f$ with respect to $\rho$ yields
\begin{equation}\label{eq:IE-Cartesian}
 \Lambda+|\vec x|^2-2\,\mathfrak{Re} f(z)-\int\!\dd^6x'\rho(\vec x')
 \log |\vec x-\vec x'|^2 \ =\ 0~,
\end{equation}
where the constraint \eqref{eq:Normalisation} is enforced by $\Lambda$. This is
the integral equation that determines the density $\rho$.  Upon acting
with $\Delta^3$ on this equation, where $\Delta$ is the Laplacian on $\IR^6$,
one quickly realises that $\rho$ cannot be an ordinary function but must be of
distributional support \cite{Berenstein:2005aa}.\footnote{In this argument, one
uses the fact that $\Delta^3_x\log|\vec x-\vec x'|\sim\delta^{(6)}(\vec x-\vec
x')$.} Thus, the integral equation \eqref{eq:IE-Cartesian}
can only hold in a suitable region of $\IR^6$. In particular, for the ground
state where $f=0$, the density $\rho$ is uniformly supported on a five-sphere
$S^5\subset\IR^6$ \cite{Berenstein:2005aa,Berenstein:2005jq}.

\begin{figure}[ht]
\hspace*{4.5cm}
\begin{picture}(240,180)
\includegraphics[width=8cm]{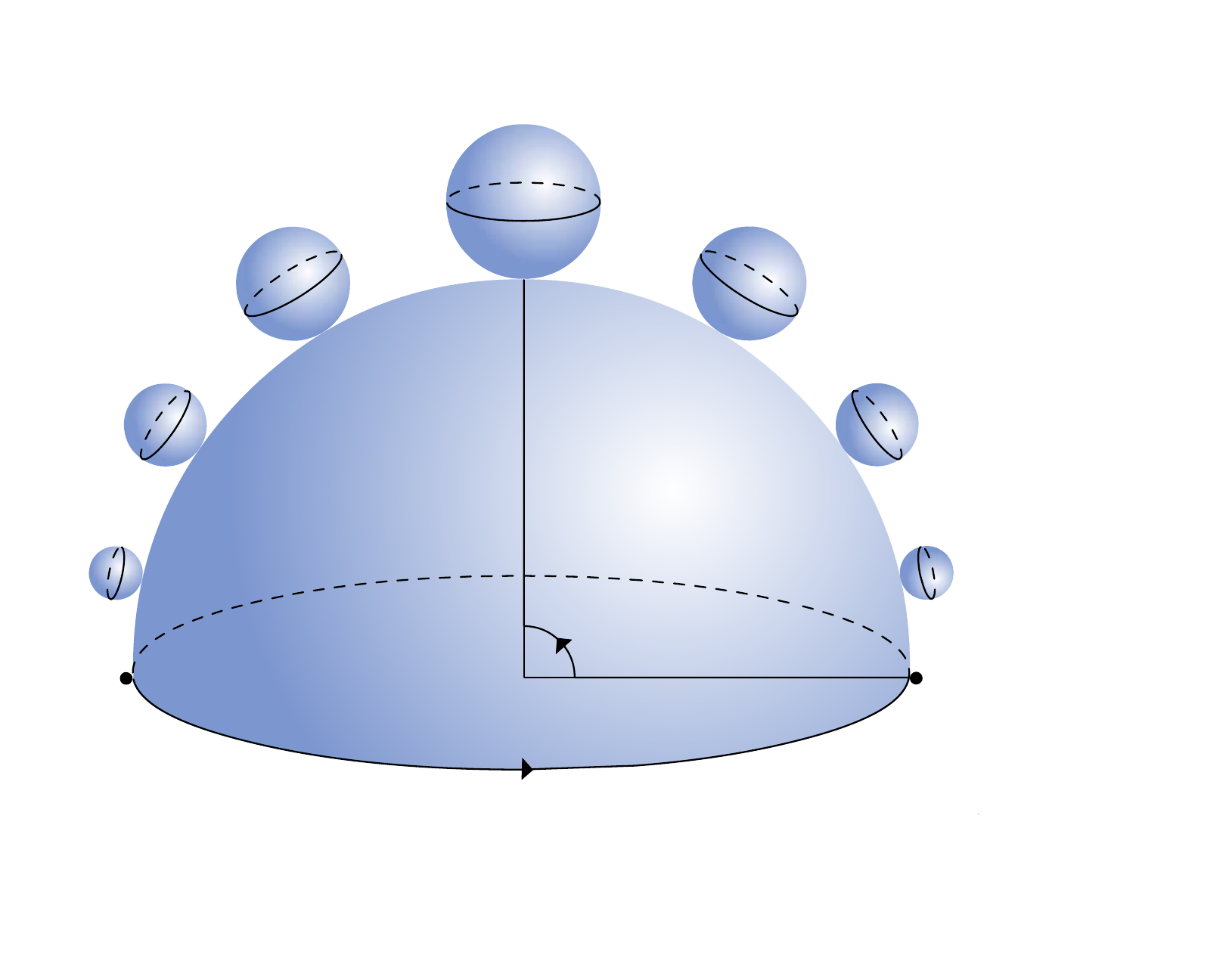}
\put(-115,65.0){\makebox(0,0)[c]{$\theta$}}
\put(-80.0,160.0){\makebox(0,0)[c]{$S^3:~(\alpha,\xi_1,\xi_2)$}}
\put(-130.0,30.0){\makebox(0,0)[c]{$\phi$}}
\end{picture}
\vspace*{-1cm}
\caption{Five-sphere described by the spherical coordinates
\eqref{eq:SphericalCoordinates} (for fixed $r$): The hemisphere is parametrised
by $\omega=(\theta,\phi)$ while the three-spheres are parametrised by
$(\alpha,\xi_1,\xi_2)$. Notice that the radii of the three-spheres shrink to
zero as $\theta$ approaches zero.}
\label{fig:FiveSphere}
\vspace*{.5cm}
\end{figure}

Instead of Cartesian coordinates $(x^1,\ldots,x^6)$, we will find it fruitful
to
make use of the following spherical parametrisation of
$\IR^6\setminus\{0\}$:
\begin{equation}\label{eq:SphericalCoordinates}
 \begin{aligned}[l]
 x^1\ =\ r \sin\theta \sin\alpha \sin\xi_1~,~~~~
  &x^2\ =\ r \sin\theta \sin\alpha \cos\xi_1~,~~~
  &x^3\ =\ r \sin\theta \cos\alpha \sin\xi_2~,\\
 x^4\ =\ r \sin\theta \cos\alpha \cos\xi_2~,~~~
  &x^5\ =\ r \cos\theta \cos\phi~,~~~
  &x^6\ =\ r \cos\theta \sin\phi~,~~~~~~~
 \end{aligned}
\end{equation}
with $\theta,\,\alpha\in[0,\pi/2]$ and $\phi,\,\xi_{1,2}\in[0,2\pi)$. Notice
that with this choice of coordinates, we have
\begin{equation}
 z\ =\ x^5+\di x^6\ =\ r\cos\theta\,\de^{\di\phi}~.
\end{equation}

To proceed we should make an ansatz for the density $\rho$. We will take it to
be supported on a hypersurface in $\IR^6$. Then, since in all the cases we
consider the potential deformation $f=f(z)$ depends only on the holomorphic
coordinates $z$, we propose the following ansatz for the density $\rho$:
\begin{equation}\label{eq:AnsatzRho}
 \rho(\vec x)\ =\ \frac{N}{2\pi^2\hat r(\theta,\phi)^5}\,\delta(r-\hat
 r(\theta,\phi))\,\hat\rho(\theta,\phi)~.
\end{equation}
Here, $\hat\rho=\hat\rho(\theta,\phi)$ and $\hat r=\hat r(\theta,\phi)$ are
non-negative functions which depend only on $\theta$ and $\phi$ and $\delta$
indicates Dirac's delta function. The problem of finding the density $\rho$ has
thus been translated into finding the functions $\hat\rho$ and $\hat r$. Notice
that the constraint \eqref{eq:Normalisation} then becomes
\begin{equation}
\int_0^{\pi/2}\!\!\!\!\dd\theta\int_0^{2\pi}\!\!\!\!\dd\phi\,
\cos\theta\sin^3\theta\,\hat \rho(\theta,\phi)\ =\ 1~.
\end{equation}

In what follows, we shall simplify notation and make use the abbreviations
$\omega:=(\theta,\phi)$ together with\\[-10pt]
\begin{equation}
\begin{minipage}{3cm}
$$
 \begin{aligned}
  \int\!\dd\Omega_2 \ &:=\
   \int_0^{\pi/2}\!\!\!\!\dd\theta\int_0^{2\pi}\!\!\!\!\dd\phi\,
   \cos\theta\sin^3\theta~, \\
  \int\!\dd\Omega_3 \ &:=\ \int_0^{\pi/2}\!\!\!\!\dd\alpha\int_0^{2\pi}
   \!\!\!\!\dd\xi_1\int_0^{2\pi}\!\!\!\!\dd\xi_2\,
   \sin\alpha\cos\alpha~,
\end{aligned}
$$
\end{minipage}
 \kern5cm
 \Longrightarrow~~~~~\int\!\dd\Omega_5\ :=\
 \int\!\dd\Omega_2\!\int\!\dd\Omega_3~.
 \vspace*{7pt}
\end{equation}

Next we wish to substitute the ansatz \eqref{eq:AnsatzRho} into
\eqref{eq:HContinuous} and derive the corresponding equations of motion.
To this end, we need the expression
\begin{equation}
 |\vec x-\vec x'|^2\,\Big|_{\!\!\scriptsize \begin{matrix}
                                   |\vec x|=\hat r(\omega)\\
                                   ~~|\vec x'|=\hat r(\omega')
                                  \end{matrix}}
  =\  \hat r^2(\omega)+\hat r^2(\omega')-2\,\hat r(\omega)\hat
  r(\omega')\cos\varphi~,
\end{equation}
where $\varphi$ is the angle $\angle\big(\frac{\vec x}{|\vec x|},\frac{\vec
x'}{|\vec x'|}\big)$ between $\frac{\vec x}{|\vec x|}$ and $\frac{\vec
x'}{|\vec
x'|}$ on the five-sphere defined by
\begin{equation}\label{eq:DefinitionAngle}
\begin{aligned}
 \cos\varphi\ &:=\ \cos\theta\cos\theta'\cos(\phi-\phi')\\
  &~~~~~~~~~~ +\,
  \sin\theta\sin\theta'\left[\cos\alpha\cos\alpha'\cos(\xi_1-\xi_1')
  +\sin\alpha\sin\alpha' \cos(\xi_2-\xi_2')\right].
\end{aligned}
\end{equation}
Therefore, we obtain for \eqref{eq:HContinuous} (including the Lagrange
multiplier term)
\begin{equation}
\begin{aligned}
 \cH_f\ &=\ \Lambda\left(\int\!\dd\Omega_2\,\hat\rho(\omega)-1\right)+N \int
 \!\dd\Omega_2\,\hat\rho(\omega)
 \left[\hat r^2(\omega) - 2\,\mathfrak{Re} f(\hat z)\right]\\
 &~~~~~~~~~~~-\,\frac{N^2}{8\pi^4}
 \int\!\dd\Omega_5\!\int\!\dd\Omega'_5\,\hat\rho(\omega)\hat\rho(\omega')
 \log\left[\hat r^2(\omega)+\hat r^2(\omega')-2\,\hat r(\omega)\hat
  r(\omega')\cos\varphi\right],
\end{aligned}
\end{equation}
where $\hat z=\hat z(\omega)=\hat r(\omega)\cos\theta\,\de^{\di\phi}$.
Variations with respect to $\Lambda,\,\hat\rho$ and $\hat r$ lead
to the following set of equations of motion:
\begin{subequations}\label{eq:IE-Spherical}
\begin{eqnarray}
 \kern-1cm 0 \!&=&\! \int\!\dd\Omega_2\,\hat\rho(\omega) - 1~,
\label{eq:IE-Spherical-l}\\
 \kern-1cm 0 \!&=&\! \frac{\Lambda}N + \hat r^2(\omega)-2\,
                     \mathfrak{Re}f(\hat z)\notag\\
 \kern-1cm      &&~~~~~~~-\,\frac{N}{4\pi^4}\int\!\dd\Omega_3
                     \!\int\!\dd\Omega'_5\,\hat\rho(\omega')
                     \log\left[\hat r^2(\omega)+\hat r^2(\omega')^2-2\,
                     \hat r(\omega)\hat r(\omega')\cos\varphi\right],
\label{eq:IE-Spherical-rho}\\
 \kern-1cm 0 \!&=&\! \hat r(\omega) -\frac{\partial\,\mathfrak{Re}
                     f(\hat z)}{\partial \hat r(\omega)}
                     -\frac{N}{4\pi^4}\int\!\dd\Omega_3\!
                     \int\!\dd\Omega'_5\,
                     \frac{\hat \rho(\omega')\left[\hat r(\omega)-
                     \hat r(\omega')\cos\varphi\right]}{\hat r^2(\omega)
                     +\hat r^2(\omega')-2\,
                     \hat r(\omega)\hat r(\omega')\cos\varphi}~.
\label{eq:IE-Spherical-r}
\end{eqnarray}
\end{subequations}
It is far from obvious how to deal with this system of integral equations in
the general case. Take notice that the equations are non-linear in $\hat r$,
and
entangle $\hat\rho$ and $\hat r$  in a  non-trivial fashion.

Nevertheless, for the ground state wavefunction where $f=0$, it is known that
the functional $\cH_f$ is extremised by \cite{Berenstein:2005jq}
\begin{equation}\label{eq:GSS}
 \Lambda\ =\ N^2\Lambda_0\ =\ N^2(\log N+\tfrac{1}{12}-\log2)~,\quad
 \hat \rho(\omega)\ =\ \rho_0\ =\ \frac2\pi~,\quad
 \hat r(\omega)\ =\ r_0\ =\ \sqrt{\frac N2}~.
\end{equation}
This exact solution will play a key role in our subsequent discussion.

\section{Analytical solutions}\label{sec:Analytical}

In this section, we will present analytical solutions to
\eqref{eq:IE-Spherical}
for $f\neq0$ and  hence, the most likely distributions of eigenvalues for
excited wavefunctions $\hat\psi_f$ (i.e.~eigenfunctions of the Hamiltonian
$H$).
We shall also compare our analytical results against numerical calculations.

\subsection{Perturbative expansions}

The starting point for our considerations is the ground state configuration
\eqref{eq:GSS}: We consider wavefunctions that can be regarded as a slight
perturbation of the ground state. For them we engineer a solution to the
equations of motion as a perturbative expansion around the ground state
solution. For this expansion to be consistent, we must assume that the function
$f=f(z)$ is `small' when evaluated on solutions to \eqref{eq:IE-Spherical},
i.e.~we introduce some small parameter $\varepsilon$ and write
\begin{equation}
 f(z)\ :=\ \varepsilon \, r_0^2  F(z)\quad\Longrightarrow\quad
 |f(z)|\ \ll\ r_0^2\ =\ \frac{N}2~,
\end{equation}
with $F$ being holomorphic in $z$. The unknowns $\Lambda$, $\hat\rho$
and $\hat r$ are then expanded in powers of $\varepsilon$ according to:
\begin{equation}\label{eq:PertSol}
 \begin{aligned}
  \Lambda \ &=\  N^2(\Lambda_0+ \varepsilon\, \Lambda_1 +\varepsilon^2\,
                \Lambda_2 + \cdots)~,\\
  \hat\rho(\omega)\ &=\ \rho_0\,( 1+ \varepsilon\, \rho_1(\omega)
                        +\varepsilon^2\, \rho_2(\omega) + \cdots)~,\\
  \hat r(\omega) \ &=\ r_0\,(1+ \varepsilon\, r_1(\omega) +\varepsilon^2\,
                      r_2(\omega)+ \cdots)~,
 \end{aligned}
\end{equation}
where $\Lambda_0$, $\rho_0$ and $r_0$ were given in \eqref{eq:GSS}.

The equations of motion \eqref{eq:IE-Spherical} can now be solved order by
order in powers of $\varepsilon$ to eventually arrive at a perturbative
solution of the form \eqref{eq:PertSol}. The upshot of this expansion is that
at any given order $k+1$ (with $k\geq0$), the integral equations for
$\rho_{k+1}=\rho_{k+1}(\omega)$ and $r_{k+1}=r_{k+1}(\omega)$ are decoupled and
linearised. Specifically, upon substituting \eqref{eq:PertSol} into
\eqref{eq:IE-Spherical}, a few algebraic manipulations show that
\begin{subequations}\label{eq:IE-Perturbative}
\begin{eqnarray}
   0 \!&=&\! \int\!\dd\Omega_2\,\rho_{k+1}(\omega)~, \label{eom1i}\\
   P_{k+1}(\omega) \!&=&\! \Lambda_{k+1}
             -\int\!\dd\Omega'_2\, K_{\rm I}(\omega,\omega')\,
             \rho_{k+1}(\omega')~,\label{eq:IE-Perturbative-rho}\\
   R_{k+1}(\omega) \!&=&\! \frac23 r_{k+1}(\omega)
      +\int\!\dd\Omega'_2\,K_{\rm II}(\omega,\omega')\,r_{k+1}(\omega')~.
   \label{eq:IE-Perturbative-r}
\end{eqnarray}
\end{subequations}
Here, we have introduced the integral kernels
\begin{equation}\label{eq:Kernels}
\begin{aligned}
 K_{\rm I}(\omega,\omega') \ &:=\ \frac{1}{2\pi^5}
    \int\!\dd\Omega_3\!\int\!\dd\Omega'_3\,\log\,(1-\cos\varphi)~,\\
 K_{\rm II}(\omega,\omega') \ &:=\ \frac{1}{2\pi^5}
    \int\!\dd\Omega_3\!\int\!\dd\Omega'_3\,\frac{1}{1-\cos\varphi}~.
\end{aligned}
\end{equation}
The remainders $P_{k+1}=P_{k+1}(\omega)$ and $R_{k+1}=R_{k+1}(\omega)$ are
functions which depend only on the solutions $\Lambda_l$, $\rho_l$ and $r_l$,
with $l\leq k$. They can be obtained order by order and the first few
expressions are
\begin{equation}\label{eq:PR-1st}
 \begin{aligned}
   P_1(\omega)\ &=\ \mathfrak{Re}F(\hat z)\big|_{\hat r=r_0}
                    +\frac{1}{\pi}\int\!\dd\Omega_2\, R_1(\omega)~,\\
   R_1(\omega)\ &=\ \hat r\frac{\partial}{\partial\hat r}
\bigg|_{\hat r=r_0} \mathfrak{Re} F(\hat z)
 \end{aligned}
\end{equation}
and
\begin{equation}\label{eq:PR-2nd}
 \begin{aligned}
  P_2(\omega) \ &=\ \frac13\,r_1^2(\omega) +\frac 1\pi
  \int\!\dd\Omega_2\,r_1(\omega)\left[2\rho_1(\omega)-r_1(\omega)\right]\\
  &~~~~~~~~~~~~~~~
  +\, \frac12\int\!\dd\Omega'_2\, K_{\rm II}(\omega,\omega')\, r_1^2(\omega')
  +\frac{1}{\pi}\int\!\dd\Omega_2\, R_2(\omega)~,\\
  R_2(\omega) \ &=\ r_1(\omega)\left(\hat
  r\frac{\partial}{\partial\hat r}\right)^2\bigg|_{\hat r=r_0}\!\!\!\!\!\!
  \mathfrak{Re}F(\hat z) -\frac{5}{3}r_1(\omega)^2+r_1(\omega)
  \int\!\dd\Omega'_2\,K_{\rm II}(\omega,\omega')\,\rho_1(\omega')\\
  &~~~~~~~~~~~~~~~
  -\,\frac12\int\!\dd\Omega'_2\,K_{\rm II}(\omega,\omega')\,
  r_1(\omega')\left[2\rho_1(\omega')-r_1(\omega')\right].
\end{aligned}
\end{equation}
Notice that upon integrating equation \eqref{eq:IE-Perturbative-rho}, the
coefficients $\Lambda_{k+1}$ for the Lagrange multiplier $\Lambda$ are given by
\begin{equation}
 \Lambda_{k+1}\ =\ \frac{2}{\pi}\int\!\dd\Omega_2\,P_{k+1}(\omega)~,
\end{equation}
since
$\int\!\dd\Omega_2\,K_I(\omega,\omega')=\int\!\dd\Omega'_2\,K_I(\omega,
\omega')=$ const and $\int\!\dd\Omega_2\,\rho_{k+1}(\omega)=0$; see Appendix
\ref{app:AppendixKernel} for more details. Hence, at any given order $k+1$, the
coefficient $\Lambda_{k+1}$ is determined by the solutions $\Lambda_l$,
$\rho_l$
and $r_l$, with $l\leq k$ and does not depend on $\rho_{k+1}$ and $r_{k+1}$.

Equations \eqref{eq:IE-Perturbative-rho} and \eqref{eq:IE-Perturbative-r}
are Fredholm integral equations of the first and second kind, respectively.
Such
kind of integral equations can in principle be solved if the eigenfunctions of
their kernels are known. We derive the eigenfunctions of
$K_{\rm I}(\omega,\omega')$ and $K_{\rm II}(\omega,\omega')$ in Appendix
\ref{app:AppendixKernel}, where we also collect some facts about these kernels.
Here we just give the results of these considerations. It happens to be that
\begin{subequations}\label{eq:Eigenfunctions}
\begin{equation}
 \Psi_{m,a}(\omega)\ =\ \Theta_{m,a}(\theta)\,\Phi_m(\phi)~,
\end{equation}
for $m\in\IZ$ and $a\in\IN_0$ with
\begin{equation}
 \Theta_{m,a}(\theta)\ :=\ \cos^{|m|}\theta\, P^{(1,|m|)}_a(\cos2\theta)\eand
 \Phi_m(\phi)\ :=\ \exp(\di m\phi)~,
\end{equation}
\end{subequations}
where $P^{(\alpha,\beta)}_a$ are the Jacobi polynomials (see Appendix
\ref{app:Jacobi}~for definitions and properties), are the eigenfunctions of
both
kernels, though with different eigenvalues:
\begin{equation}
 \int\!\dd\Omega'_2\, K_{\rm I,II}(\omega,\omega')\, \Psi_{m,a}(\omega')\ =\
 \lambda_{\rm I,II}^{m,a}\,\Psi_{m,a}(\omega)~,
\end{equation}
with
\begin{equation}\label{eq:Eigenvalues}
\begin{aligned}
 \lambda_{\rm I}^{m,a}\ &= \
    \begin{cases}
     {\displaystyle \frac{7}{12}-\log2}&\efor(m,a)\ =\ (0,0)~,\\[5pt]
     {\displaystyle -24\,\frac{(|m|+2a-1)!}{(|m|+2a+4)!}}
           &\efor(m,a)\ \neq\ (0,0)~,\\
    \end{cases}\\[5pt]
  \lambda_{\rm II}^{m,a}\ &=\ 8\,\frac{(|m|+2a)!}{(|m|+2a+3)!}~.
\end{aligned}
\end{equation}
Notice that the functions \eqref{eq:Eigenfunctions} form a complete orthogonal
basis for functions defined on the hemisphere given by $\omega=(\theta,\phi)$;
see Figure \ref{fig:FiveSphere}.

Therefore, we may expand the remainders appearing in \eqref{eq:IE-Perturbative}
in terms of \eqref{eq:Eigenfunctions},
\begin{equation}
 P_{k+1}(\omega)\ =\ \sum_{m,a}P_{k+1}^{m,a}\,\Psi_{m,a}(\omega)\eand
 R_{k+1}(\omega)\ =\ \sum_{m,a}R_{k+1}^{m,a}\,\Psi_{m,a}(\omega)~.
 \vspace*{-5pt}
\end{equation}
Since $P_{k+1}$ is real, we must have $P_{k+1}^{m,a}=(P_{k+1}^{-m,a})^*$ and
similarly for $R_{k+1}$; `$*$' indicates complex conjugation. Likewise,
$\Lambda_{k+1}$, $\rho_{k+1}$ and $r_{k+1}$
can be expanded in terms of \eqref{eq:Eigenfunctions}. Upon replacing all of
these expansions in the integral equations \eqref{eq:IE-Perturbative}, we find
that $\Lambda_{k+1}$, $\rho_{k+1}$ and $r_{k+1}$ are given by
\begin{subequations}\label{eq:PerturbativeSolution}
\begin{eqnarray}
 \Lambda_{k+1} \!&=&\! P_{k+1}^{0,0}~,\\[10pt]
 \rho_{k+1}(\omega)\!&=&\!-\!\!\!\!\!\sum_{(m,a)\neq(0,0)}
 \frac{P_{k+1}^{m,a}}{\lambda_{\rm I}^{m,a}}\,\Psi_{m,a}(\omega)~,
 \label{eq:PertSolrho}\\
 r_{k+1}(\omega)\!&=&\! \sum_{m,a} \frac{ R_{k+1}^{m,a}}{\tfrac{2}3+
 \lambda_{\rm II}^{m,a}}\,\Psi_{m,a}(\omega)~.\label{eq:PertSolr}
\end{eqnarray}
\end{subequations}
It should be stressed that in general there is no guarantee that the
expressions
\eqref{eq:PertSolrho} and \eqref{eq:PertSolr} would define square-integrable
functions. For that to happen, the functions $P_{k+1}$ and $R_{k+1}$ should
meet
certain criteria. In particular, \eqref{eq:IE-Perturbative-rho} is a Fredholm
integral equation of the first kind with symmetric kernel and a complete set of
eigenfunctions. Therefore, certain existence theorems apply.
Specifically, \eqref{eq:IE-Perturbative-rho} has a unique $L_2$-solution given
by \eqref{eq:PertSolrho} if and only if the infinite series
\begin{equation}\label{eq:criterium}
 \sum_{(m,a)\neq(0,0)} \left|\frac{P_{k+1}^{m,a}}{\lambda_{\rm
I}^{m,a}}\right|^2
\end{equation}
is convergent. For more details, see e.g.~\cite{Tricomi:1957}.

In summary, provided certain criteria are met, we are able to solve the saddle
point equations \eqref{eq:IE-Spherical} analytically by means of the
perturbative expansions \eqref{eq:PertSol}.

\subsection{Example A: Monomial deformations}

Let us now consider an example and take $f(z)\propto z^p$ for $p\in\IN$.
Firstly, these are the simplest distortions of the ground state wavefunction.
Secondly, and more interestingly, these wavefunctions should correspond to
simple gravitational duals. Indeed, they are expected to be in correspondence
with the LLM geometries obtained from a simply connected droplet possessing a
non-vanishing $p^{\rm th}$ harmonic moment \cite{Vazquez:2006id}.

For this to constitute a small perturbation, $f(z)$ should be of order
$\varepsilon\, r_0^2$, with $\varepsilon\ll1$. Thus, the constant of the
monomial is taken to scale with $N$. In particular, we consider
\begin{equation}
 f(z)\ =\ \varepsilon\, r_0^2
\left(\frac{z}{r_0}\right)^p\quad\Longrightarrow\quad F(z)\ =\
\left(\frac{z}{r_0}\right)^p\eand
 \mathfrak{Re}F(z)\ =\ \left(\frac{r}{r_0}\right)^p\cos^p\theta\cos p\phi~.
\end{equation}
Notice that since $\mathfrak{Re}F(z)$ is an even function in $\phi$ and since
$K_{I,II}(\theta,\phi,\theta',\phi')=K_{I,II}(\theta,-\phi,\theta',-\phi')$ as
a direct consequence of \eqref{eq:DefinitionAngle} and \eqref{eq:Kernels},
the functions $\hat \rho$ and $\hat r$ will be even functions in $\phi$. Hence,
the series expansions \eqref{eq:PerturbativeSolution} reduce to expansions in
terms of $\mathfrak{Re}\Psi_{m,a}$ with $m\geq0$ and real coefficients (i.e.~we
only need the $\cos m\phi$ terms in the Fourier expansion).

Therefore, the expressions \eqref{eq:PR-1st} for the remainders $P_1$ and $R_1$
are given by
\begin{equation}
\begin{aligned}
 P_1(\omega)\ &=\ \cos^p\theta \cos p\phi\ =\
 \mathfrak{Re}\Psi_{p,0}(\omega)~,\\
 R_1(\omega)\ &=\ p\cos^p\theta \cos p\phi\ =\
p\,\mathfrak{Re}\Psi_{p,0}(\omega)~.
\end{aligned}
\end{equation}
Using these results and the series expansions \eqref{eq:PerturbativeSolution},
we then obtain
\begin{subequations}\label{eq:Solution-1st}
\begin{eqnarray}
   \Lambda_{1} \!&=&\! 0~,\\[6pt]
\rho_{1}(\omega)\!&=&\!
\frac{1}{24}\frac{(p+4)!}{(p-1)!}\,\cos^p\theta \cos p\phi~,\\
   r_{1}(\omega)\!&=&\!
 \frac{3 p (p+3)!}{2(p+3)!+24 p!}\,\cos^p\theta \cos p\phi~.
\end{eqnarray}
\end{subequations}

\begin{table}[h]
\centering
\begin{tabular}{cccccccc}
\hline\hline\\[-10pt]
$p$ & $\rho_2^{2p,0}$ & $\rho_2^{0,0}$ & $\rho_2^{0,1}$ & $\rho_2^{0,2}$ &
$\rho_2^{0,3}$ & $\rho_2^{0,4}$ \\[3pt]
\hline\\[-10pt]
2 & $\frac{925}{3}$ & 0 & 10 & $\frac{185}6$ & $0$ & $0$  \\[5pt]
3 & $\frac{435375}{121}$ & 0 & $\frac{2430}{121}$ & $\frac{89910}{847}$ &
$\frac{87075}{847}$ & $0$  \\[5pt]
4 & $\frac{29458800}{1369}$ & 0 & $\frac{40320}{1369}$ & $\frac{7560}{37}$ &
$\frac{481600}{1369}$ & $\frac{233800}{1369}$ \\[8pt]
\hline
\end{tabular}
\caption{Coefficients of $\rho_2(\omega)$.}
\label{tab:Coeff-rho}
\end{table}

\begin{table}[h]
\centering
\begin{tabular}{cccccccc}
\hline\hline\\[-10pt]
$p$ & $r_2^{2p,0}$ & $r_2^{0,0}$ & $r_2^{0,1}$ & $r_2^{0,2}$ & $r_2^{0,3}$ &
$r_2^{0,4}$  \\[3pt]
\hline\\[-10pt]
2 & $\frac{2875}{592}$ & $-\frac{115}{32}$ & 0 & $\frac{575}{1184}$ & $0$ & $0$
 \\[5pt]
3 & $\frac{42795}{1892}$ & $-\frac{12681}{968}$ &
$-\frac{1125}{484}$ & $\frac{492075}{250712}$ & $\frac{8559}{13244}$ & $0$
\\[5pt]
4 & $\frac{14391825}{228623}$ & $-\frac{136885}{4107}$ & $-\frac{43840}{4107}$
&
$\frac{348705}{101306}$ & $\frac{477680}{176601}$ & $\frac{685325}{1371738}$
\\[8pt]
\hline
\end{tabular}
\caption{Coefficients of $r_2(\omega)$.}
\label{tab:Coeff-r}
\end{table}

Next we would like to compute the solution to order $\varepsilon^2$. As it
is clear from \eqref{eq:PR-2nd} and \eqref{eq:Solution-1st}, in order to
compute
$\Lambda_2$, $\rho_2$ and $r_2$ we need to expand $(\cos^p\theta\cos
p\phi)^2=[\mathfrak{Re}\Psi_{p,0}(\omega)]^2$ in terms of the eigenfunctions
\eqref{eq:Eigenfunctions}. We find
\begin{equation}
 [\mathfrak{Re}\Psi_{p,0}(\omega)]^2\ =\
  \frac12\, \mathfrak{Re}\Psi_{2p,0}(\omega)+\sum_{q=0}^p
  \frac{(q+1)^2 p!^2}{(p-q)!(p+q+2)!}\, \mathfrak{Re}\Psi_{0,q}(\omega)~.
\end{equation}
Therefore, using \eqref{eq:PR-2nd}, the non-vanishing coefficients $P_2^{m,a}$
and $R_2^{m,a}$,
\begin{equation}
 P_{2}(\omega)\ =\
 \sum_{m,a}P_{2}^{m,a}\,\mathfrak{Re}\Psi_{m,a}(\omega)\eand
 R_{2}(\omega)\ =\ \sum_{m,a}R_{2}^{m,a}\,\mathfrak{Re}\Psi_{m,a}(\omega)~,
 \vspace*{-5pt}
\end{equation}
are computed to be
\begin{subequations}
\begin{equation}
\begin{aligned}
 \!\!\! P_2^{m,a}\ &=\ \begin{cases}
 {\displaystyle\frac{b_p c_{p,0}}{2}\!\left[p^2+\left(\frac23+
 \lambda_{\rm II}^{p,0}\right)a_p\right]}&\!\!\!\!\!\!\efor\! (m,a)\ =\
 (0,0)~,\\[5pt]
 {\displaystyle\frac{b_p^2}{4}\!\left(\frac23+
 \lambda_{\rm II}^{2p,0}\right)}&\!\!\!\!\!\!\efor\! (m,a)\ =\
 (2p,0)~,\\[5pt]
 {\displaystyle\frac{b_p^2c_{p,q}}{2}\!\left(\frac23+
 \lambda_{\rm II}^{0,q}\right)}
 &\!\!\!\!\!\!\efor\! (m,a)\ =\ (0,1\leq q\leq p)~,\end{cases}\\
 \!\!\! R_2^{m,a}\ &=\ \begin{cases}
 {\displaystyle\frac{b_p}{2}\!\left[p^2-\frac53\,b_p+a_p
 \lambda_{\rm II}^{p,0}-\left(a_p-\frac{b_p}{2}\right)\!
 \lambda_{\rm II}^{2p,0}\right]}&\!\!\!\!\!\!\efor\! (m,a)\ =\ (2p,0)~,\\[5pt]
 {\displaystyle b_pc_{p,q}\!\left[p^2-\frac53\,b_p+a_p\lambda_{\rm II}^{p,0}-
 \left(a_p-\frac{b_p}{2}\right)\!\lambda_{\rm II}^{0,q}\right]}
 &\!\!\!\!\!\!\efor\! (m,a)\ =\ (0,0\leq q\leq p)~,
 \end{cases}
\end{aligned}
\end{equation}
where
\begin{equation}
 a_p\ :=\ \frac{1}{24}\frac{(p+4)!}{(p-1)!}~,\quad
 b_p\ :=\ \frac{3 p (p+3)!}{2(p+3)!+24 p!}\eand
 c_{p,q}\ :=\ \frac{(q+1)^2 p!^2}{(p-q)!(p+q+2)!}~
\end{equation}
\end{subequations}
and the eigenvalues $\lambda_{\rm I,II}^{m,a}$ were given in
\eqref{eq:Eigenvalues}. From these expressions, we obtain
\begin{equation}
 \Lambda_2\ =\ P_2^{0,0}~,\quad
 \rho_2^{m,a}\ =\ -\frac{P_{2}^{m,a}}{\lambda_{\rm I}^{m,a}}\efor (m,a)\ \neq\
 (0,0) \eand r_2^{m,a}\ =\ \frac{ R_{2}^{m,a}}{\tfrac{2}3+
 \lambda_{\rm II}^{m,a}}~
\end{equation}
as the only non-vanishing coefficients appearing in
\begin{equation}
 \rho_{2}(\omega)\ =\
 \sum_{m,a}\rho_{2}^{m,a}\,\mathfrak{Re}\Psi_{m,a}(\omega)\eand
 r_{2}(\omega)\ =\ \sum_{m,a}r_{2}^{m,a}\,\mathfrak{Re}\Psi_{m,a}(\omega)~.
 \vspace*{-5pt}
\end{equation}
We refrain from writing down these expressions explicitly, as they are lengthy
and not particularly illuminating. Instead, we only list the particular values
for $p=2,3,4$ in Tables \ref{tab:Coeff-rho} and  \ref{tab:Coeff-r}.

\begin{figure}[h!]
\hspace*{.1cm}
\begin{picture}(200,140)
\includegraphics[scale=0.45]{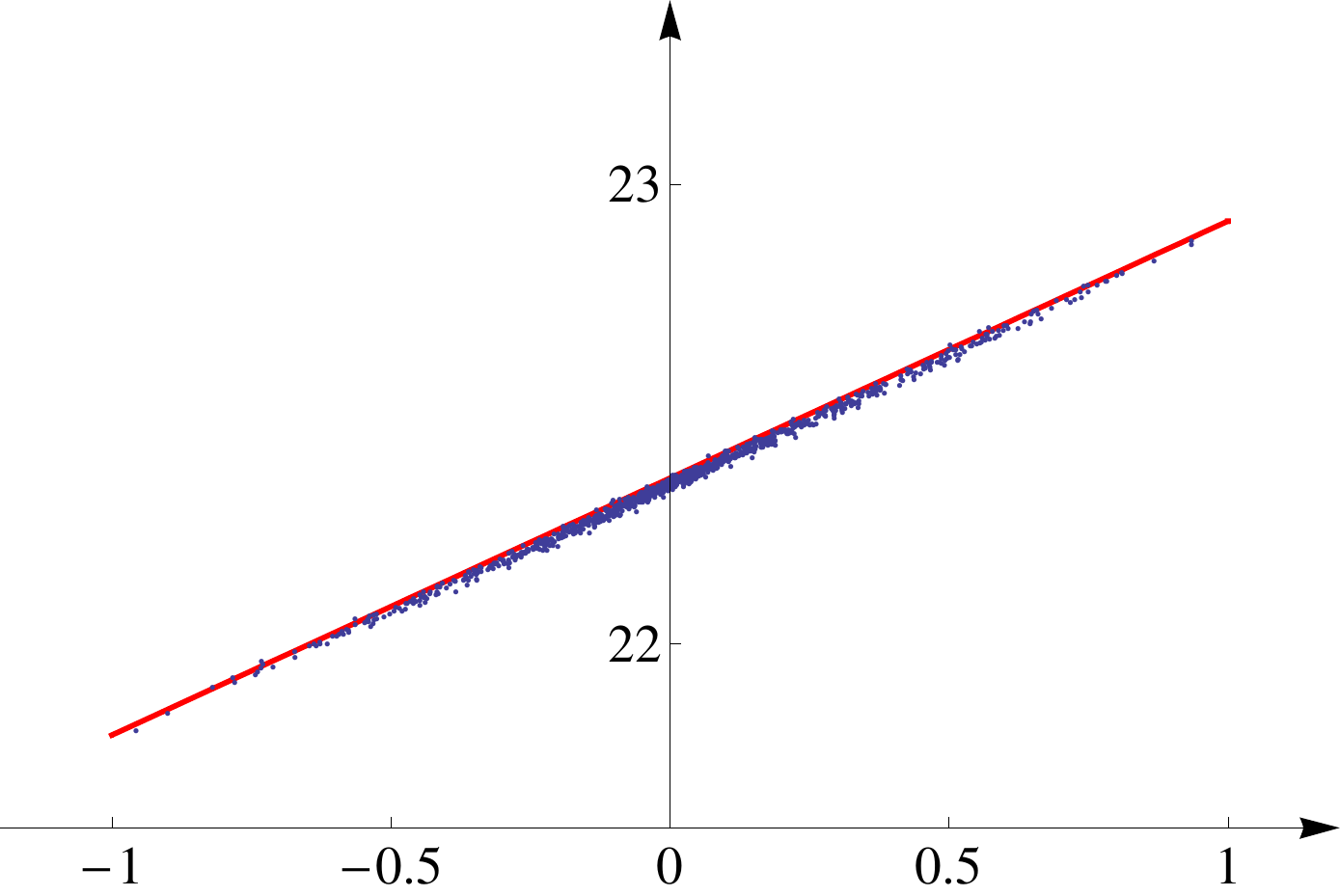}
\put(-90.0,128.0){\makebox(0,0)[c]{$^{\hat r}$}}
\put(28.0,8.0){\makebox(0,0)[c]{$^{\cos^2\theta\cos 2\phi}$}}
\put(-90.0,-8.0){\makebox(0,0)[c]{\small (a)}}
\put(-30.0,40.0){\makebox(0,0)[c]{\tiny {\cor $22.34 -0.558 \cos^2\theta\cos
2\phi$}}}
\put(70.0,60.0){\makebox(0,0)[l]{\begin{minipage}{7cm}
\underline{Case $p=2$:}\\  This is very close to the analytical solution
$\sqrt{500}(1-\tfrac{115}{32}\varepsilon^2 - \tfrac52 \varepsilon
\cos^2\theta\cos 2\phi) \approx 22.35 - 0.559 \cos^2\theta\cos 2\phi$.
\end{minipage}}}
\end{picture}

\vspace*{15pt}
\hspace*{.1cm}
\begin{picture}(200,140)
\includegraphics[scale=0.45]{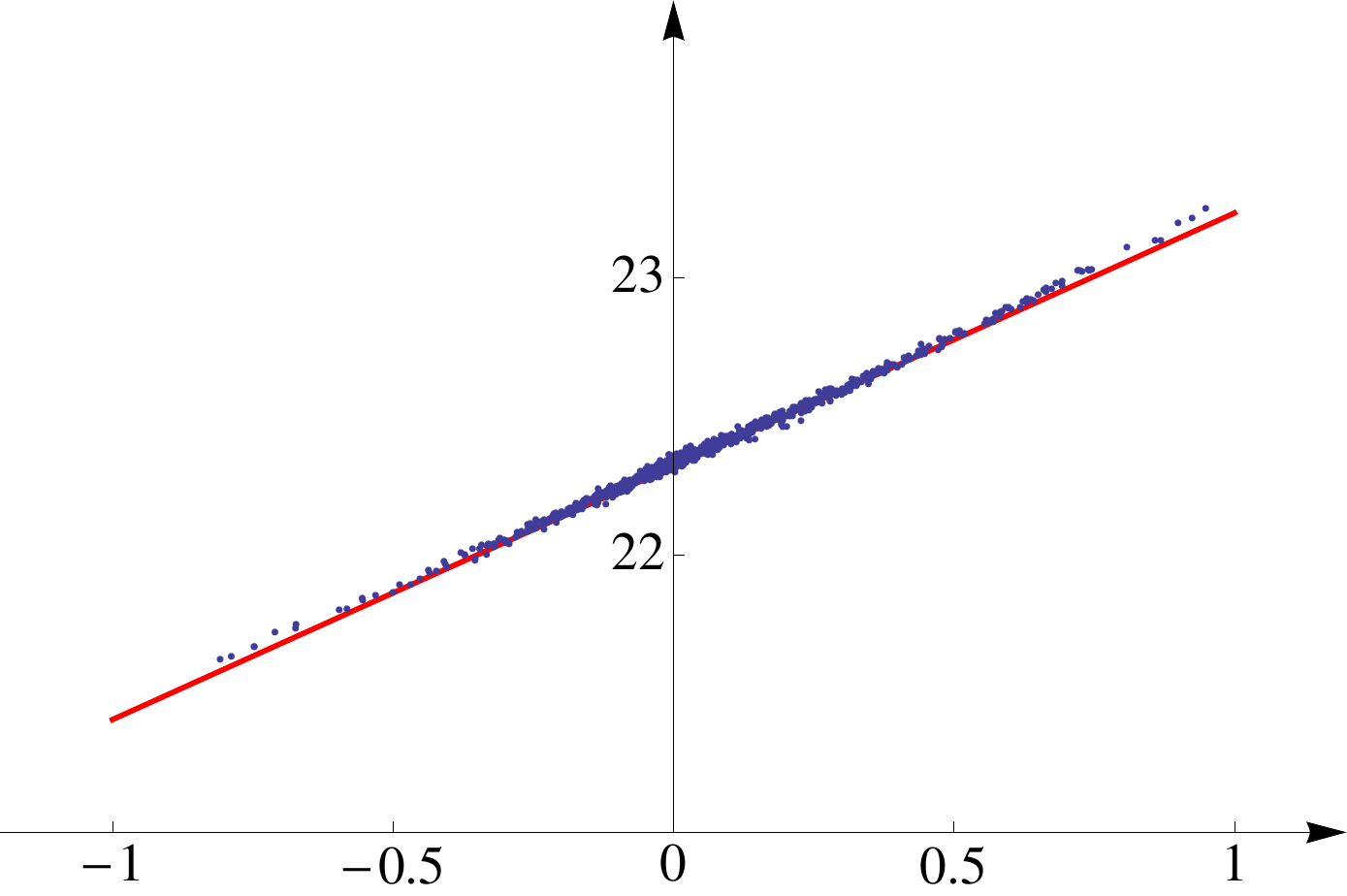}
\put(-90.0,128.0){\makebox(0,0)[c]{$^{\hat r}$}}
\put(28.0,8.0){\makebox(0,0)[c]{$^{\cos^3\theta\cos 3\phi}$}}
\put(-90.0,-8.0){\makebox(0,0)[c]{\small (b)}}
\put(-30.0,40.0){\makebox(0,0)[c]{\tiny {\cor $22.32 -0.915 \cos^3\theta\cos
3\phi$}}}
\put(70.0,60.0){\makebox(0,0)[l]{\begin{minipage}{7cm}
\underline{Case $p=3$:}\\ This is very close to the analytical solution
$\sqrt{500}(1-\tfrac{12681}{968}\varepsilon^2 - \tfrac{45}{11} \varepsilon
\cos^3\theta\cos 3\phi) \approx 22.33 - 0.915 \cos^3\theta\cos
3\phi$.\end{minipage}}}
\end{picture}

\vspace*{15pt}
\hspace*{.1cm}
\begin{picture}(200,140)
\includegraphics[scale=0.45]{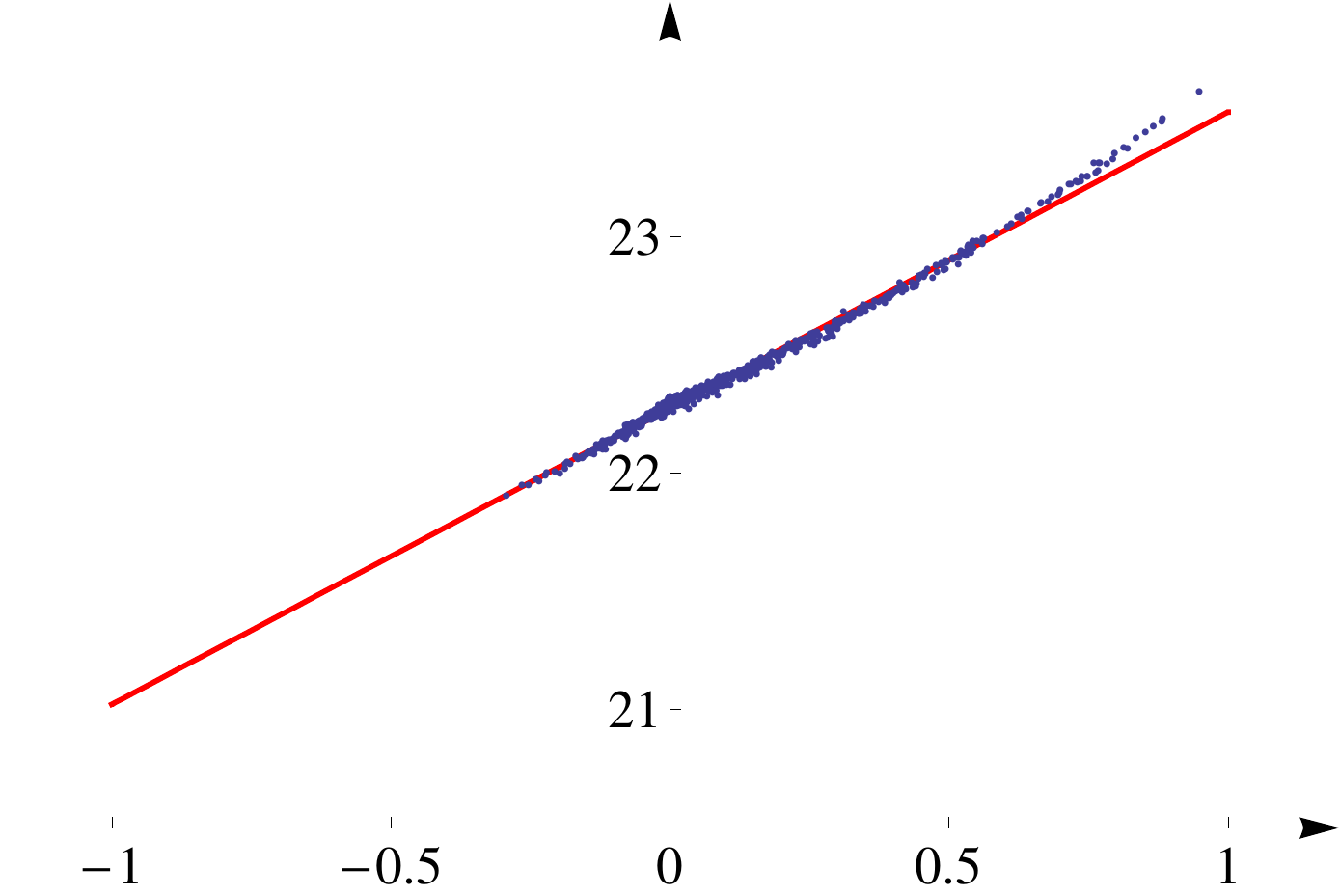}
\put(-90.0,128.0){\makebox(0,0)[c]{$^{\hat r}$}}
\put(28.0,8.0){\makebox(0,0)[c]{$^{\cos^4\theta\cos 4\phi}$}}
\put(-90.0,-8.0){\makebox(0,0)[c]{\small (b)}}
\put(-30.0,40.0){\makebox(0,0)[c]{\tiny {\cor $22.27 -1.253 \cos^4\theta\cos
4\phi$}}}
\put(70.0,60.0){\makebox(0,0)[l]{\begin{minipage}{7cm}
\underline{Case $p=4$:}\\ This is still close to the analytical solution
$\sqrt{500}(1-\tfrac{136885}{4107}\varepsilon^2 - \tfrac{210}{37} \varepsilon
\cos^4\theta\cos 4\phi) \approx 22.29 - 1.269 \cos^4\theta\cos 4\phi$. The
slight deviation is due to the fact that for larger $p$ the numerical value of
the (neglected) higher-order coefficients is larger.
\end{minipage}}}
\end{picture}

\vspace*{10pt}
\caption{Numerical solutions for $p=2,3,4$ and $\varepsilon=0.01$ obtained with
$N=1000$. The solid curves represent linear fittings of the data points.}
\label{fig:Monomial}
\end{figure}

In what follows, in order to compare against the analytical results, we will
compute these coefficients out of numerical solutions. To have a numerical
handle on the problem, we can retrace our steps from the continuum limit to
finite $N$. For finite $N$, the extrema of $\cH_f$ will not dominate the
partition function \eqref{eq:partitionfunction} anymore as the saddle point
approximation is only valid in the thermodynamic limit $N\to\infty$. Instead,
for a finite $N$ description, one should take into account all possible
configurations weighted by their respective probability. In order to do that in
a good approximation, Monte Carlo simulations with a Metropolis criterion can
be
used to simulate the distribution of probabilities: Starting off from a random
distribution of particles, a new one is generated by a small random
perturbation. The new configuration is accepted if $\de^{-\delta\cH_f}$ (where
$\cH_f$ is the discrete Hamiltonian \eqref{eq:Hdiscrete}) is larger than a
random number in the interval $[0,1]$ and rejected otherwise. By iterating this
algorithm a large number of times, a typical (i.e.~most likely) configuration
of
particles will be obtained at the end. This method was used to describe
different wavefunctions of the matrix model we are dealing with for finite $N$
in works by Berenstein  and collaborators
\cite{Berenstein:2007wz,Berenstein:2008jn,Berenstein:2010dg}.

At this point, we should emphasise that the remainder of this section is not
aimed at finding the finite $N$ description of wavefunctions. Instead, we will
use the finite $N$ problem as a discretisation of the equations of motions of
the continuum limit, which, of course, would be valid in the $N \to \infty$
limit. In other words, we will consider the finite $N$ Hamiltonians $\cH_f$ and
still look for their extrema numerically and regard these as numerical
approximations of the perturbative analytical solutions presented previously.
To
do that, we will simply modify the Metropolis criterion to accept new
configurations only if $\delta{\cH_f} <0$. In this way, the iteration procedure
will produce configurations approaching the extrema of $\cH_f$. All the
simulations in this work that use this modified Metropolis criterion were
performed in {\ttfamily Mathematica}.

Using that method we obtained numerical approximations for the cases $p=2,3,4$
with $\varepsilon=0.01$. In these three cases, we discretised the densities
using $N=1000$. For such small $\varepsilon$, the main dependence of $\hat r$
will be that of the linear order in $\varepsilon$. If one plots $\hat r$ versus
$\cos^p\theta\cos p\phi$, the points should approximately lie in a line. In
Figure \ref{fig:Monomial} we produce those plots on top of their linear
fittings. The agreement of these numerical fittings with the analytical
coefficients is very good. Certainly, beyond the the leading approximation,
$\hat r$ is not a function of $\cos^p\theta\cos p\phi$ only. By fitting $\hat
r$
as an expansion of the eigenfunctions kicking in at higher orders in
$\varepsilon$ one can obtain the corresponding numerical coefficients.
For instance, for $p=2$, we get the coefficients
\begin{subequations}
\begin{equation}
 (r_2^{4,0},r_2^{0,0},r_2^{0,1},r_2^{0,2})\big|_{\mbox{\tiny numerical}}\
\approx\ (5.1,
-3.7 , 0.02 , 0.58)
\end{equation}
which are fairly close to the coefficients in first row of Table
\ref{tab:Coeff-r}:
\begin{equation}
 (r_2^{4,0},r_2^{0,0},r_2^{0,1},r_2^{0,2})\big|_{\mbox{\tiny analytical}}\ =\
 (\tfrac{2875}{592},-\tfrac{115}{32},0,\tfrac{575}{1184})\ \approx\
 (4.7,-3.6,0,0.49)~.
\end{equation}
\end{subequations}

It is also possible to compute the density coefficients for the numerical
solutions using\footnote{ The $\rho^{m,a}$s are the coefficients of the
full $\hat\rho$,
i.e.~$\hat\rho(\omega)=\sum_{m,a}\rho^{m,a}\mathfrak{Re}\Psi_{m,a}(\omega)$.}
\begin{subequations}
\begin{equation}
 \rho^{m,a} =  \frac{1}{\CN_{m,a}}\int\!\dd\Omega_2\,\hat\rho(\omega)\,
 \mathfrak{Re}\Psi_{m,a}(\omega)\ \to\
 \frac{1}{\CN_{m,a}}\,\frac{1}{N}\sum_{i=1}^N
\mathfrak{Re}\Psi_{m,a}(\omega_i)~,
\end{equation}
with the normalisation constant
\begin{equation}
 \CN_{m,a}\ :=\ \frac{\pi}{2}\frac{(a+1)(1+\delta_{m0})}{(2a+m+2)(a+m+1)}~,
\end{equation}
\end{subequations}
where $\delta$ is the Kronecker delta. For $p=2$, we obtain
\begin{subequations}
\begin{equation}
 (\rho_1^{2,0},\rho_2^{4,0},\rho_2^{0,1},\rho_2^{0,2})\big|_{\mbox{\tiny
numerical}}\
\approx\ (30.1,306,10.1,30.0)
\end{equation}
in good agreement with the coefficients in Table \ref{tab:Coeff-rho}:
\begin{equation}
 (\rho_1^{2,0},\rho_2^{4,0},\rho_2^{0,1},\rho_2^{0,2})\big|_{\mbox{\tiny
analytical}}\ =\ (30,\tfrac{925}{3},10,\tfrac{185}{6})\approx\
(30,308.3,10,30.8)~.
\end{equation}
\end{subequations}

\subsection{Example B: Logarithmic deformations}

Let us now consider a wavefunction with a logarithmic potential $f(z)= Q \log
z$. In this case, the wavefunction should correspond to the BPS operator $\det
Z^Q$ \cite{Berenstein:2005aa}. In turn, this operator is believed to be dual
to the annular LLM geometry whose droplet inner and outer radii  are $\sqrt{Q}$
and $\sqrt{N+Q}$, respectively.

One could expect this case to be simpler because the $\phi$-dependence drops
out
of the problem (notice that the real part of $f$ is independent of $\phi$) and
the functions $\hat r$ and $\hat \rho$ will only depend on $\theta$. On the
other hand, the distribution cannot be regarded as a small distortion of the
spherical for all values of the angle $\theta$ (since $f$ has singularities)
and
therefore we cannot use \eqref{eq:PertSol} to solve the equations
perturbatively.

Nevertheless, there is a simplification since for a logarithmic potential, the
term $\frac{\partial\,\mathfrak{Re} f}{\partial\hat r}$ appearing in the
`radial' integral equation \eqref{eq:IE-Spherical-r} does not depend on
$\theta$
explicitly. Therefore, \eqref{eq:IE-Spherical-r} can be solved exactly by
assuming a constant radius $\hat r$:
\begin{equation}\label{eq:theorad}
 \hat r\ =\ r_Q\ =\ \sqrt{\frac N2+Q}~.
\end{equation}
To arrive at this result, we have used the constraint \eqref{eq:IE-Spherical-l}
to eliminate $\hat\rho$. We should emphasise that this solution is an exact
solution for \emph{any} value of $Q$.

\begin{figure}[h!]
\centering
\begin{picture}(160,170)
\includegraphics[scale=0.4]{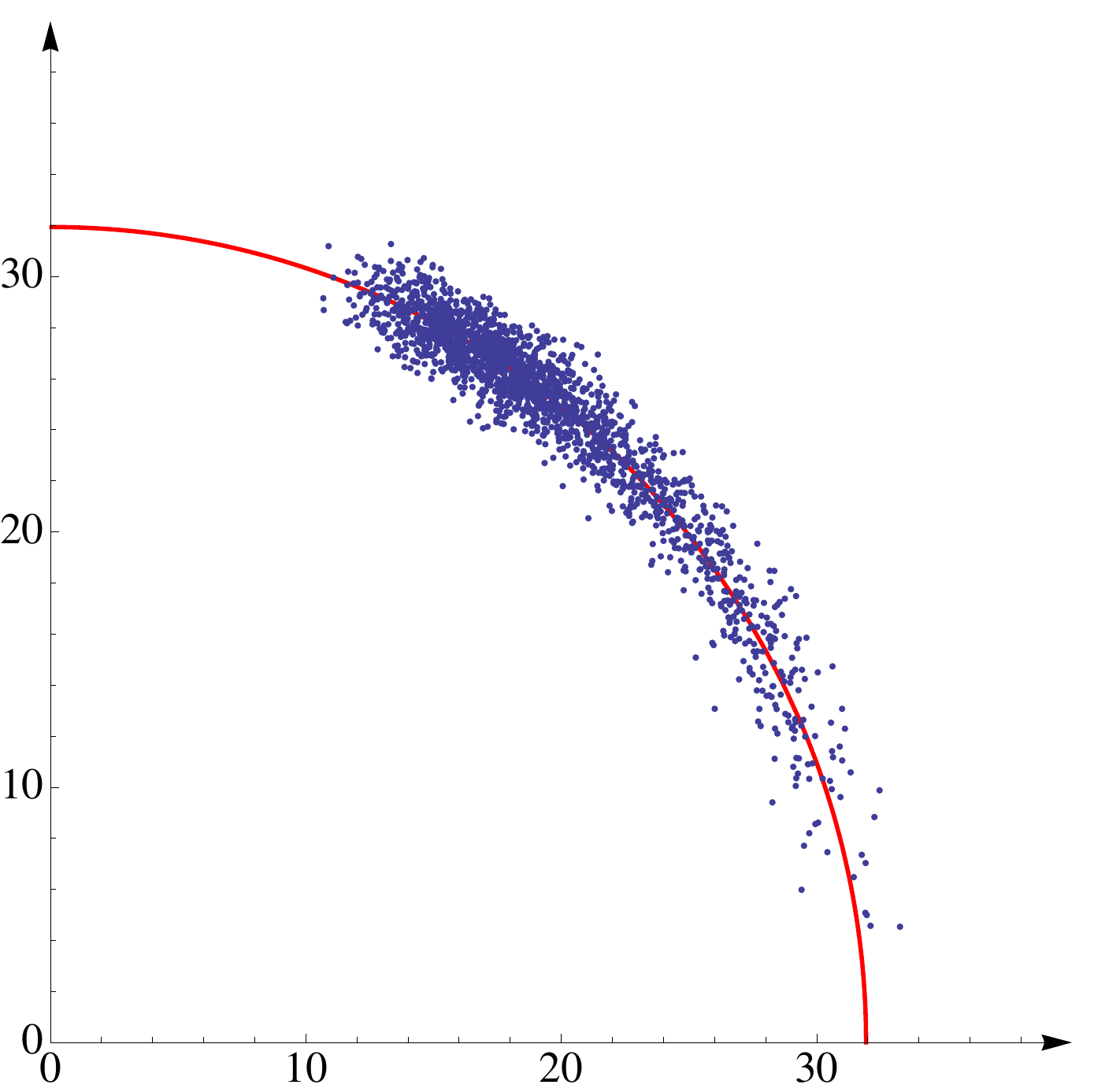}
\put(-153.0,162.0){\makebox(0,0)[c]{$^{R_{56}}$}}
\put(10.0,5.0){\makebox(0,0)[c]{$^{R_{1234}}$}}
\put(-80,-10.0){\makebox(0,0)[l]{\small (a)}}
\end{picture}
\hspace*{1.5cm}
\begin{picture}(160,170)
\includegraphics[scale=0.4]{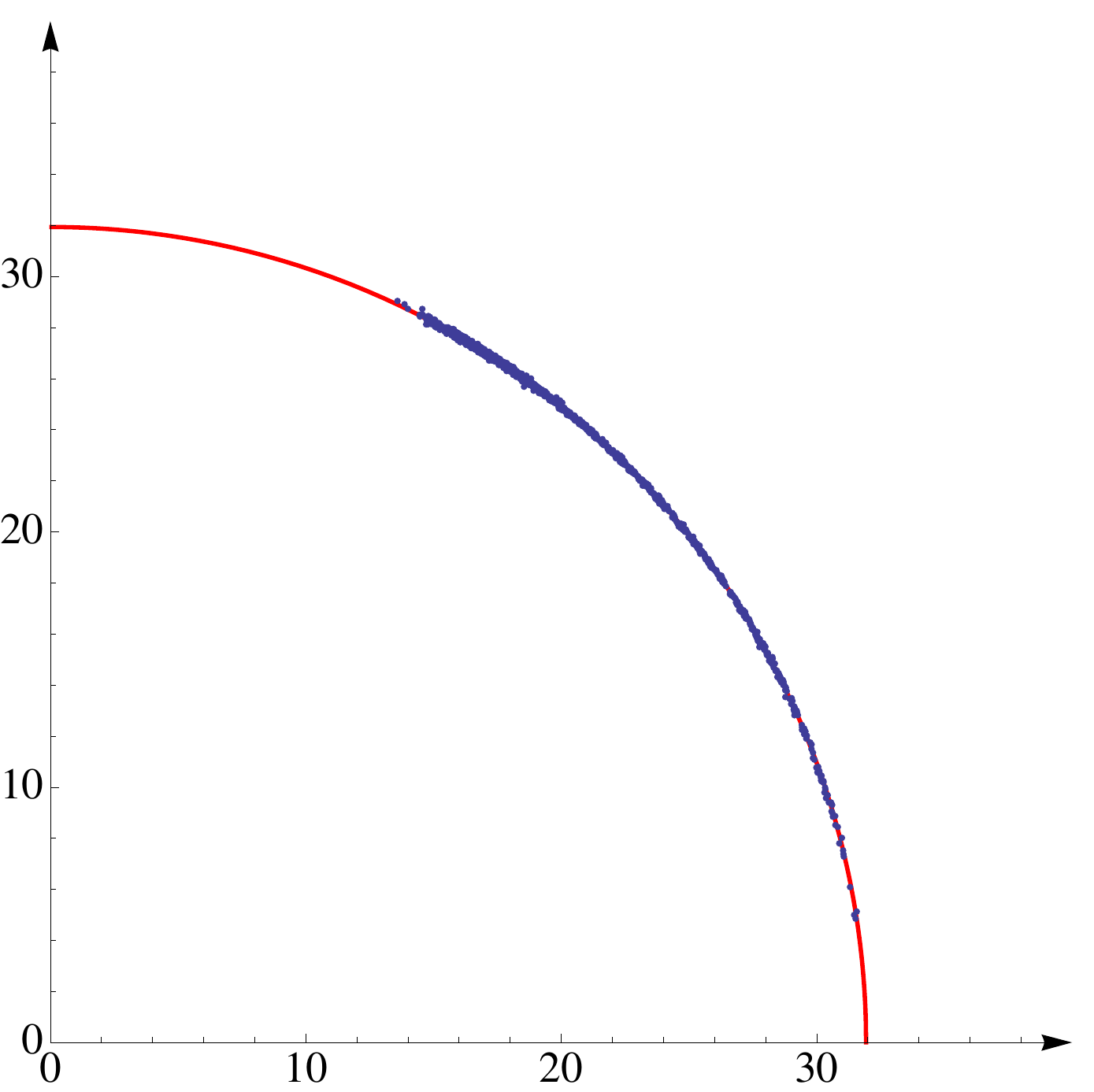}
\put(-153.0,162.0){\makebox(0,0)[c]{$^{R_{56}}$}}
\put(10.0,5.0){\makebox(0,0)[c]{$^{R_{1234}}$}}
\put(-80,-10.0){\makebox(0,0)[l]{\small (b)}}
\end{picture}
\vspace*{10pt}
\caption{On the left, a typical configuration for $N=2000$
and $Q=20$ obtained by using a Metropolis criterion in the numerical
simulation.
On the right, a configuration extremising $\cH_f$ which is also obtained for
$N=2000$ and $Q=20$. In both pictures, the solid curve is
$R_{1234}^2+R_{56}^2=1020$.}
\label{fig:logx2}
\end{figure}
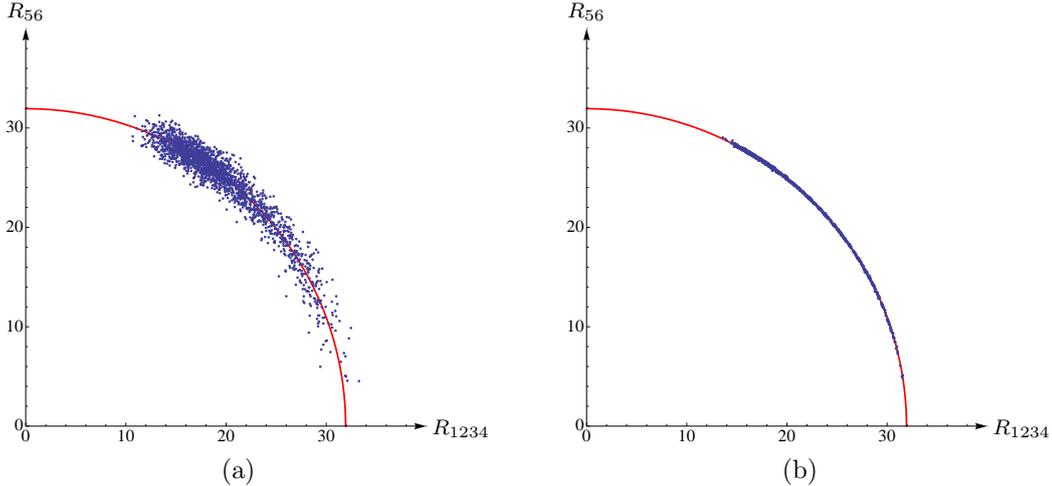

Pleasingly, this appears to be consistent with Monte Carlo simulations for
finite values of $N$ \cite{Berenstein:2007wz,Berenstein:2010dg}. In Figure
\ref{fig:logx2}, we have depicted $R_{56}:=\sqrt{x_5^2+x_6^2}$ versus
$R_{1234}:=\sqrt{x_1^2+x_2^2+x_3^2+x_4^2}$ for $N=2000$ particles and $Q=20$.
The plot on the left corresponds to a typical (i.e.~most likely) configuration
obtained from a Metropolis algorithm (as done in
\cite{Berenstein:2007wz,Berenstein:2010dg}). The solid curve corresponds to
 $R_{1234}^2+R_{56}^2=r_Q^2$ with $r_Q^2=1020$. Of course, the agreement
would improve for simulations with larger $N$. As before, one can modify the
Metropolis criterion to accept new configurations only if $\delta{\cH_f} <0$.
In
this way, the iteration will produce configurations approaching the extrema of
${\cH_f}$. As shown in right plot in Figure \ref{fig:logx2}, these
configurations fit very well the large-$N$ value of the constant radius
\eqref{eq:theorad}.\footnote{In fact, several simulations for
different $N$ and $Q$ values show that the particles of extremal
configurations sit at $r_Q=\sqrt{\tfrac {N-1}2+Q}$.}

Now that we have solved exactly the radial integral equation with
\eqref{eq:theorad}, we are left with a  linear integral equation for the
density
$\hat\rho(\theta)$,
\begin{subequations}
\begin{equation}\label{eq:eom-log-case}
 \tilde\Lambda-\frac{Q}{N}\log(1+\cos2\theta)
-\frac{\pi}2\int_0^{\pi/2} \!\!\dd\theta'\cos\theta'\sin^3\theta'\,
 \tilde K_I(\theta,\theta')\,\hat\rho(\theta')\ =\ 0~,
\end{equation}
with
\begin{equation}\label{eq:eom-log-case-2}
 \tilde\Lambda\ :=\ \frac{\Lambda}{N^2}+
 \frac{r^2_Q}{N}-\frac{Q}{N}\log\frac{r^2_Q}{2}-
           \log2r^2_Q\eand
 \tilde K_I(\theta,\theta')\ :=\ \int_0^{2\pi}\!\!\dd\phi\,
K_I(\omega,\omega')~.
\end{equation}
\end{subequations}
Using the expansion \eqref{eq:KernelEFExpansion} of the kernel $K_I$, we may
re-write the kernel $\tilde K_I$ as an infinite series
\begin{equation}\label{eq:TildeK}
 \tilde K_I(\theta,\theta')\ =\
4\sum_{a=0}^\infty\lambda_I^{0,a}(1+a)P_a^{(1,0)}(\cos 2\theta)P_a^{(1,0)}(\cos
2\theta')~,
\end{equation}
where $\lambda_I^{0,a}$ was given in \eqref{eq:Eigenvalues}. Likewise, we could
integrate \eqref{eq:Kernels-V4} to arrive at an infinite series in terms of
hypergeometric functions.

Equation \eqref{eq:eom-log-case} is again a Fredholm integral of the first
kind.
In general, the range of validity of such an equation does not need to coincide
with the interval of integration. In fact, \eqref{eq:eom-log-case} cannot hold
for all $0\leq \theta \leq \tfrac\pi 2$ as can be seen as
follows:\footnote{To arrive at this conclusion, one may also argue differently.
Upon expanding $\log(1+\cos2\theta)$ and $\hat\rho$ in terms of the Jacobi
polynomials $P_a^{(1,0)}(\cos 2\theta)$ and using  \eqref{eq:TildeK}, one
quickly realises that the obtained series $\hat\rho(\theta)=\sum_a\rho_a
P_a^{(1,0)}(\cos 2\theta)$ is not convergent as the coefficients $\rho_a$ do
not
satisfy the criterion \eqref{eq:criterium}. Hence, no solution exists
satisfying
the equation for all $0\leq \theta \leq \tfrac\pi 2$.} Firstly,
\eqref{eq:TildeK} implies that $\tilde K_I$ is bounded\footnote{Alternatively,
one may deduce this directly from the expressions \eqref{eq:Kernels-V4}.} since
with $|P_a^{(1,0)}(\cos 2\theta)|\leq1+a$ for $0\leq \theta \leq \tfrac\pi 2$
we
find
\begin{equation}
 |\tilde K_I(\theta,\theta')|\ \leq\ 4\sum_{a=0}^\infty
| \lambda_I^{0,a}|(1+a)^3\ =\ 4\log2~.
\end{equation}
Secondly, the integral (recall that $\hat\rho\geq0$)
\begin{eqnarray}\label{eq:Estimate}
 \kern-1cm\left| \int_0^{\pi/2} \!\!\dd\theta'\cos\theta'\sin^3\theta'\,
 \tilde K_I(\theta,\theta')\,\hat\rho(\theta')\right|\! &\leq&\!
  \int_0^{\pi/2} \!\!\dd\theta'\cos\theta'\sin^3\theta'\,
 |\tilde  K_I(\theta,\theta')|\,\hat\rho(\theta')\notag\\
  &\leq&\! 4\log 2\int_0^{\pi/2} \!\!\dd\theta'\cos\theta'\sin^3\theta'\,
  \hat\rho(\theta')\ =\ \frac{2}{\pi}\log 2~
\end{eqnarray}
is also bounded for all $0\leq\theta\leq\tfrac{\pi}{2}$. In the last step of
this derivation, we have used the normalisation of the density $\hat\rho$.
However, the logarithmic term $\log(1+\cos 2\theta)$ appearing in
\eqref{eq:eom-log-case} is not bounded. Therefore, we conclude that
\eqref{eq:eom-log-case} cannot be solved for all $0\leq \theta \leq \tfrac\pi
2$. Note, however, that this does not contradict the intuition that a function
$\hat\rho$ extremising $\cH_f$ should exist. Rather, since $\hat\rho\geq 0$,
the
extrema of $\cH_f$ could lie in the \emph{boundary} of the space of the allowed
configurations $\hat\rho$. Indeed, the above analysis indicates that the
extrema
of $\cH_f$ must be found in the boundary of the configuration space, i.e.~the
density $\hat\rho$ has to vanish in some region $U\subset[0,\tfrac\pi2]$ and in
this region $U$, the integral equation will not hold.

This is consistent with the numerical simulations, where $\hat\rho$ appears to
be vanishing in the region $U_{\theta_0}=\{\theta\,|\,\theta_0\leq \theta \leq
\tfrac\pi 2\}$ for some $\theta_0$. Altogether, we are left with the following
(one-dimensional) integral equation for $\hat\rho$:
\begin{equation}\label{eq:eom-log-case-theta0}
\tilde\Lambda-\frac{Q}{N}\log(1+\cos2\theta)
-\frac{\pi}2\int_0^{\theta_0} \!\!\dd\theta'\cos\theta'\sin^3\theta'\,
 \tilde K_I(\theta,\theta')\,\hat\rho(\theta')\ =\ 0~,
 \efor 0\ \leq\ \theta\ \leq\ \theta_0
 \end{equation}
for some value $\theta_0=\theta_0\big(\frac{Q}{N}\big)$ which one has to
determine consistently together with $\hat\rho$.\footnote{Notice that
$\theta_0\big(\frac{Q}{N}=0\big)=\tfrac\pi2$. Furthermore, a necessary
condition
is that $|\tilde\Lambda-\frac{Q}{N}\log(1+\cos2\theta_0)|\leq \log2$ as follows
from \eqref{eq:Estimate}.} To find $\theta_0$ and $\hat\rho$, one should solve
\eqref{eq:eom-log-case-theta0} for generic $\theta_0$ (not for all $\theta_0$ a
solution should be found). Among those solutions, one then should look for the
one that extremises $\cH_f$.

Unfortunately, solving the Fredholm integral equation
\eqref{eq:eom-log-case-theta0} for arbitrary $\theta_0$ is a very difficult
problem. So far, we have not been able solve the eigenvalue problem of the
kernel \eqref{eq:eom-log-case-2} for any other value than
$\theta_0=\tfrac\pi2$. For that reason, we shall determine $\hat\rho$
numerically in the following.

\begin{figure}[h!]
\centering
\hspace*{-1cm}
\begin{picture}(150,120)
\includegraphics[scale=0.4]{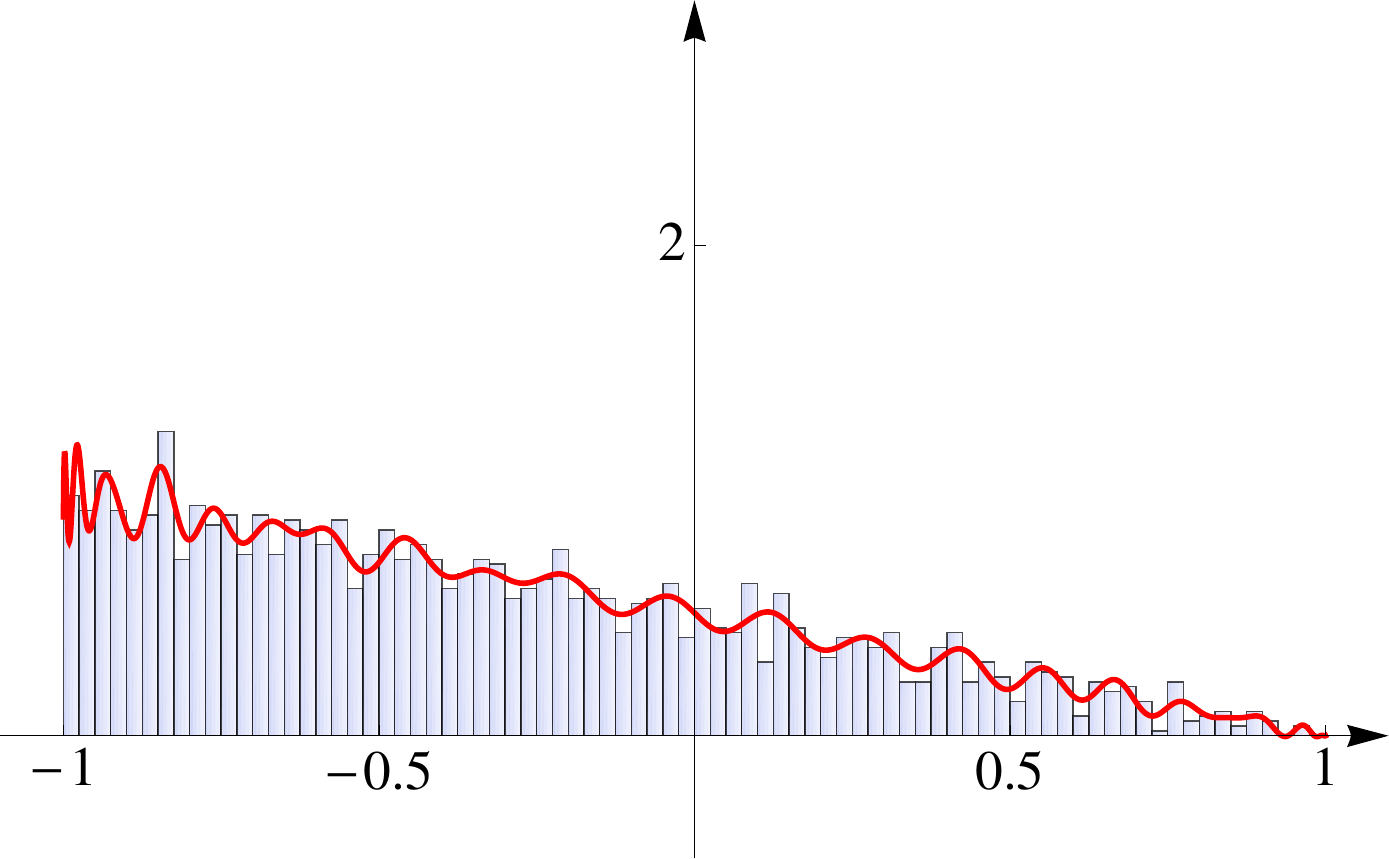}
\put(-75.0,105.0){\makebox(0,0)[c]{$^{(1-x)\hat\rho(x)}$}}
\put(8.0,5.0){\makebox(0,0)[c]{$^x$}}
\put(-60.0,80.0){\makebox(0,0)[l]{$^{\frac{Q}{N}\,=\,0}$}}
\put(-57.0,60.0){\makebox(0,0)[l]{$^{\tilde\Lambda\,\approx\,-0.1098}$}}
\put(-59.5,40.0){\makebox(0,0)[l]{$^{x_0\,\approx\,-0.9967}$}}
\put(-85,-10.0){\makebox(0,0)[l]{\small (a)}}
\end{picture}
\hspace*{2cm}
\begin{picture}(150,120)
\includegraphics[scale=0.4]{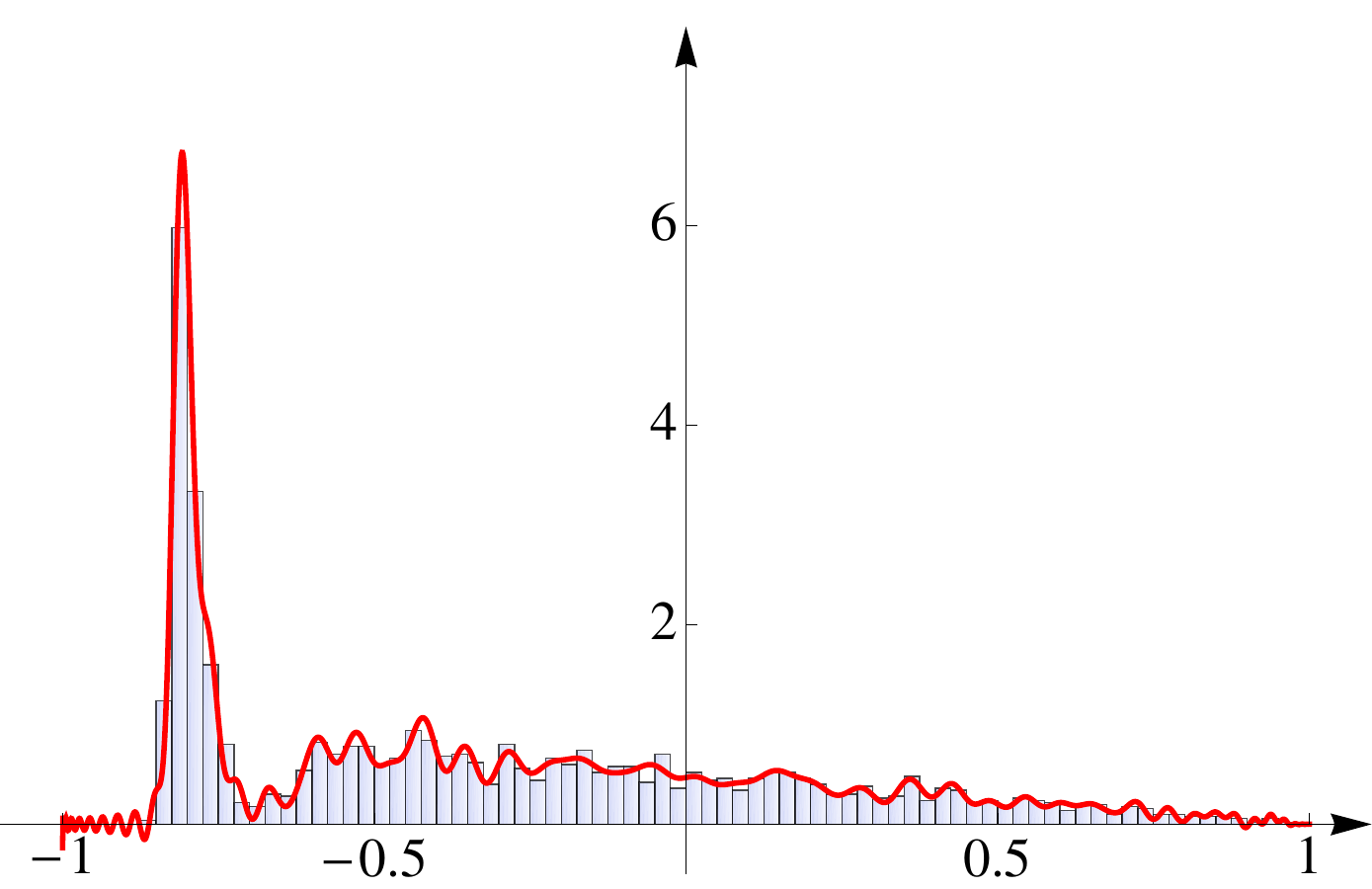}
\put(-75.0,105.0){\makebox(0,0)[c]{$^{(1-x)\hat\rho(x)}$}}
\put(8.0,5.0){\makebox(0,0)[c]{$^x$}}
\put(-60.0,80.0){\makebox(0,0)[l]{$^{\frac{Q}{N}\,=\,\frac{1}{500\pi}\,\approx\,
0.0006}$}}
\put(-57.0,60.0){\makebox(0,0)[l]{$^{\tilde\Lambda\,\approx\,-0.1103}$}}
\put(-59.5,40.0){\makebox(0,0)[l]{$^{x_0\,\approx\,-0.8495}$}}
\put(-85,-10.0){\makebox(0,0)[l]{\small (b)}}
\end{picture}

\vspace*{20pt}
\hspace*{-1cm}
\begin{picture}(150,120)
\includegraphics[scale=0.4]{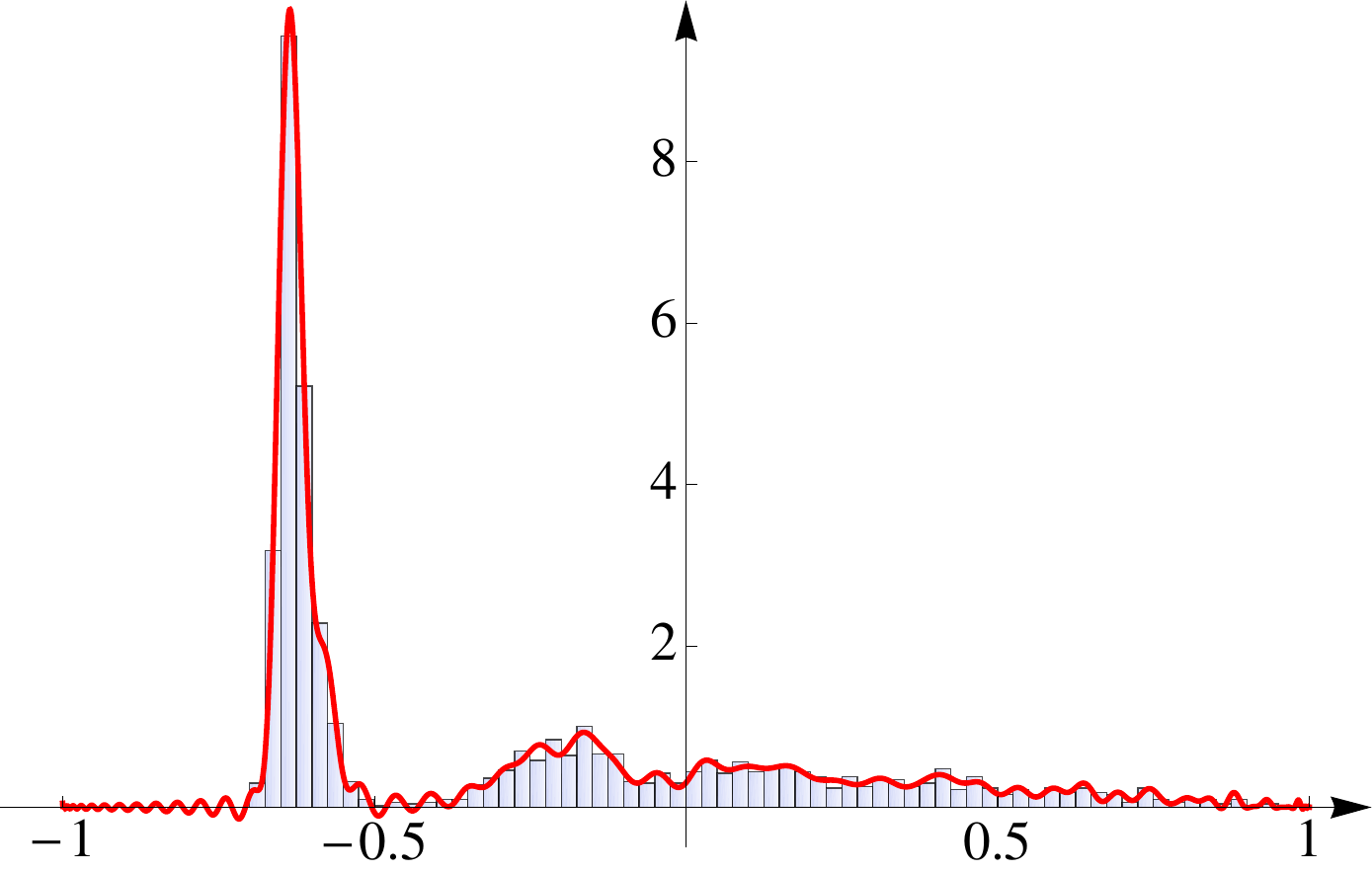}
\put(-75.0,105.0){\makebox(0,0)[c]{$^{(1-x)\hat\rho(x)}$}}
\put(8.0,5.0){\makebox(0,0)[c]{$^x$}}
\put(-60.0,80.0){\makebox(0,0)[l]{$^{\frac{Q}{N}\,=\,\frac{1}{100\pi}\,\approx\,
0.0032}$}}
\put(-57.0,60.0){\makebox(0,0)[l]{$^{\tilde\Lambda\,\approx\,-0.112}$}}
\put(-59.5,40.0){\makebox(0,0)[l]{$^{x_0\,\approx\,-0.6824}$}}
\put(-85,-10.0){\makebox(0,0)[l]{\small (c)}}
\end{picture}
\hspace*{2cm}
\begin{picture}(150,120)
\includegraphics[scale=0.4]{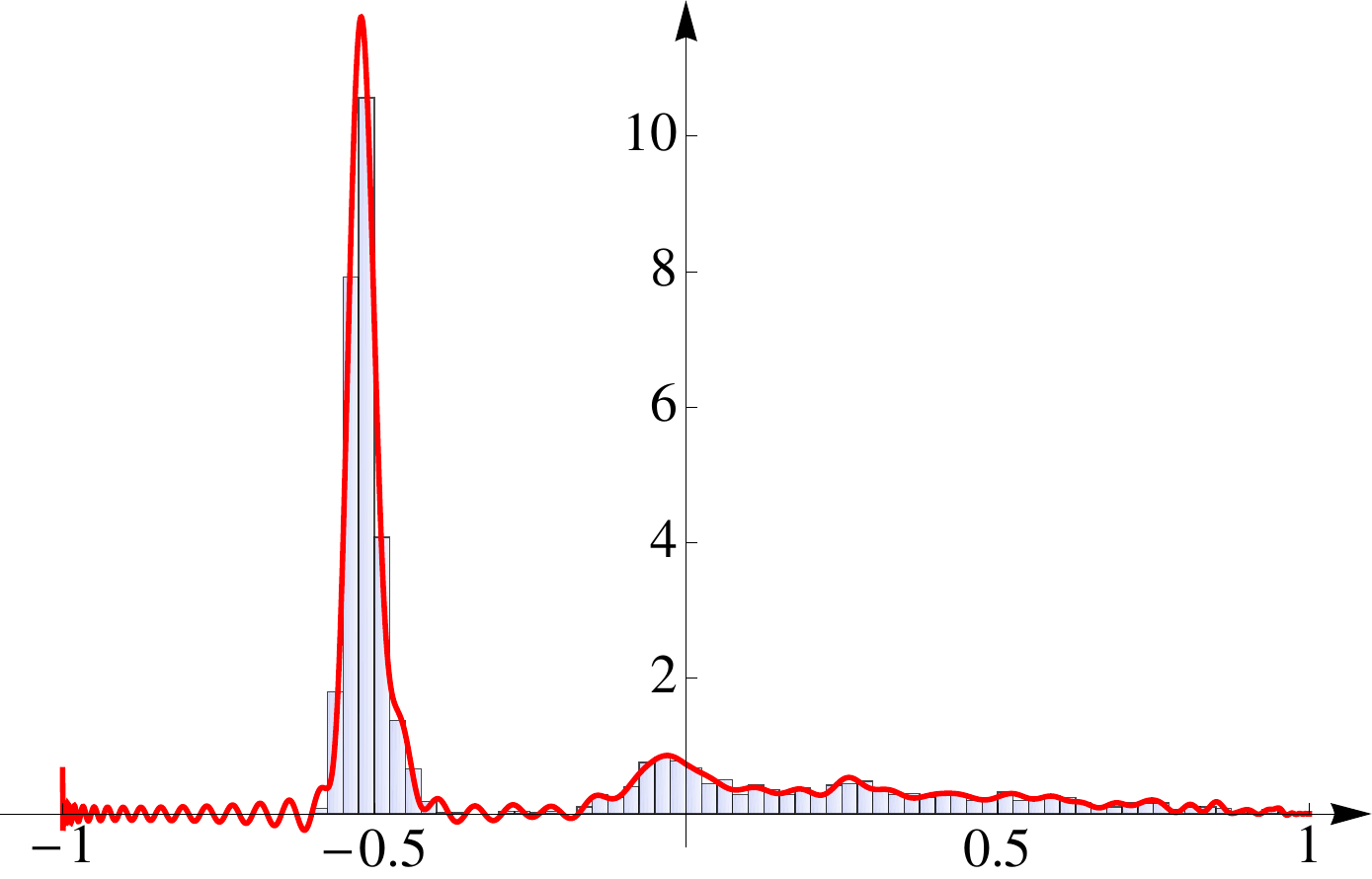}
\put(-75.0,105.0){\makebox(0,0)[c]{$^{(1-x)\hat\rho(x)}$}}
\put(8.0,5.0){\makebox(0,0)[c]{$^x$}}
\put(-60.0,80.0){\makebox(0,0)[l]{$^{\frac{Q}{N}\,=\,\frac{1}{50\pi}\,\approx\,
0.0064}$}}
\put(-57.0,60.0){\makebox(0,0)[l]{$^{\tilde\Lambda\,\approx\,-0.1139}$}}
\put(-59.5,40.0){\makebox(0,0)[l]{$^{x_0\,\approx\,-0.5803}$}}
\put(-85,-10.0){\makebox(0,0)[l]{\small (d)}}
\end{picture}

\vspace*{20pt}
\hspace*{-1cm}
\begin{picture}(150,120)
\includegraphics[scale=0.4]{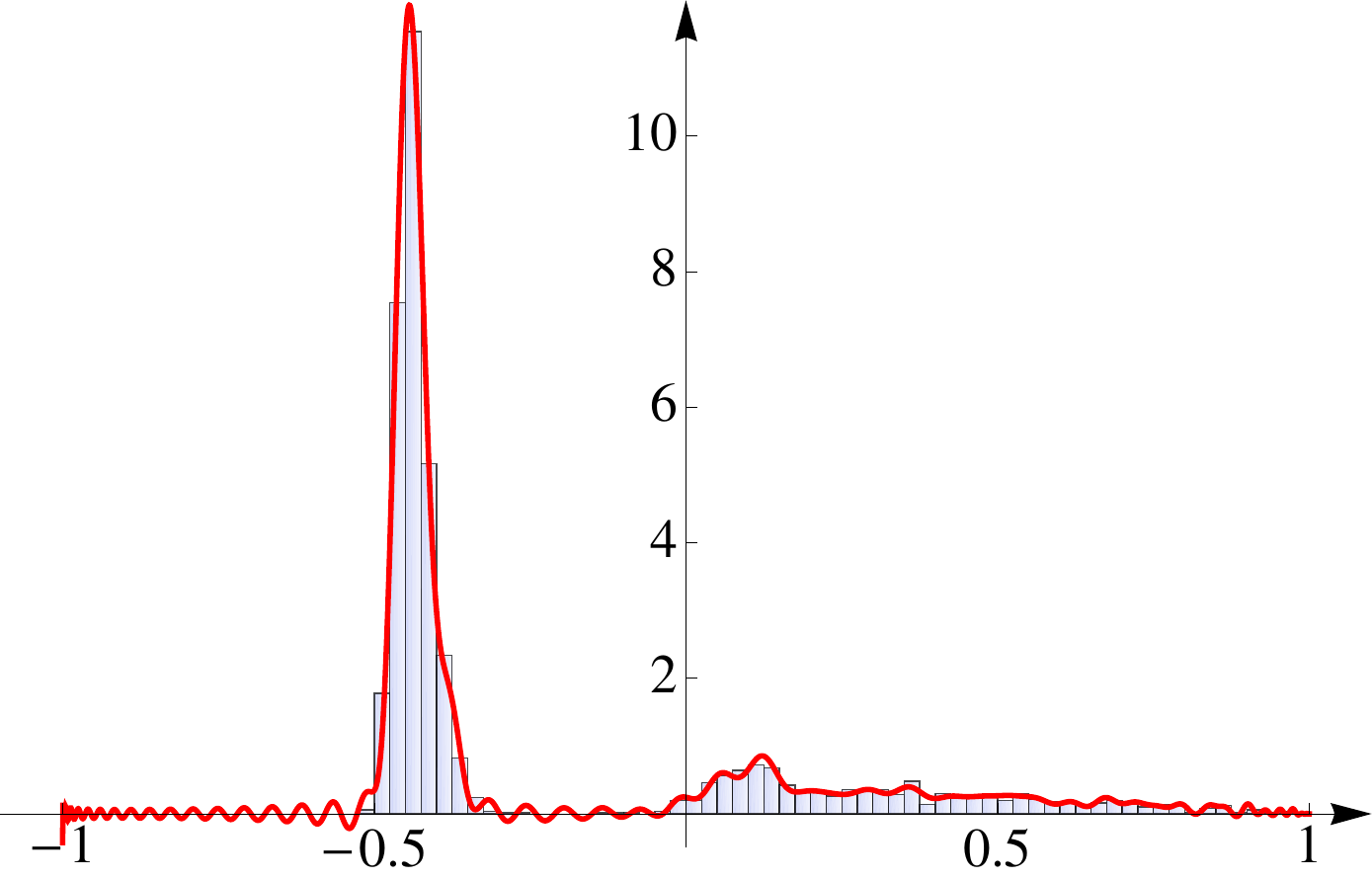}
\put(-75.0,105.0){\makebox(0,0)[c]{$^{(1-x)\hat\rho(x)}$}}
\put(8.0,5.0){\makebox(0,0)[c]{$^x$}}
\put(-60.0,80.0){\makebox(0,0)[l]{$^{\frac{Q}{N}\,=\,\frac{3}{100\pi}\,\approx\,
0.0095}$}}
\put(-57.0,60.0){\makebox(0,0)[l]{$^{\tilde\Lambda\,\approx\,-0.1157}$}}
\put(-59.5,40.0){\makebox(0,0)[l]{$^{x_0\,\approx\,-0.5024}$}}
\put(-85,-10.0){\makebox(0,0)[l]{\small (e)}}
\end{picture}
\hspace*{2cm}
\begin{picture}(150,120)
\includegraphics[scale=0.4]{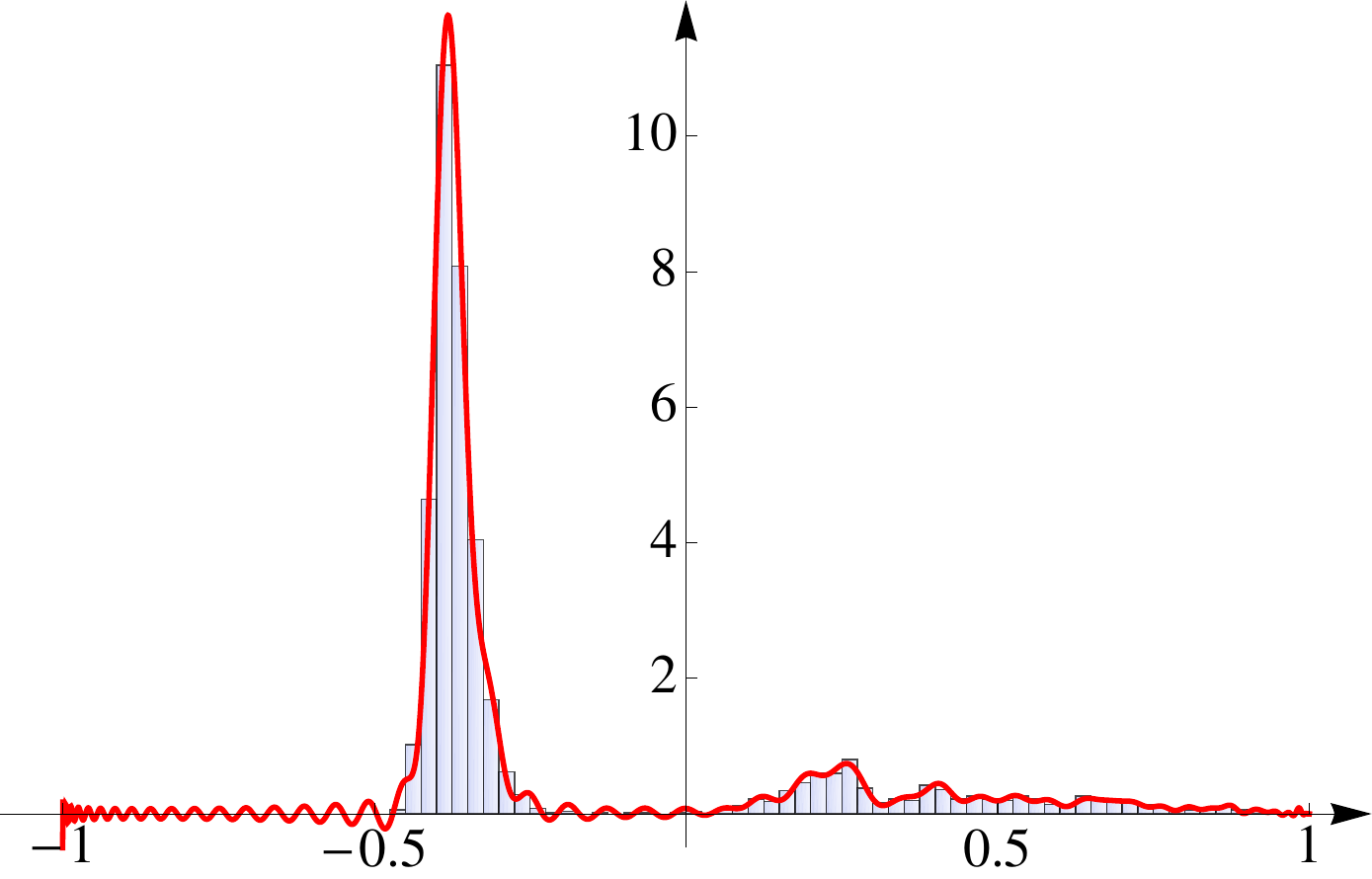}
\put(-75.0,105.0){\makebox(0,0)[c]{$^{(1-x)\hat\rho(x)}$}}
\put(8.0,5.0){\makebox(0,0)[c]{$^x$}}
\put(-60.0,80.0){\makebox(0,0)[l]{$^{\frac{Q}{N}\,=\,\frac{1}{25\pi}\,\approx\,
0.0127}$}}
\put(-57.0,60.0){\makebox(0,0)[l]{$^{\tilde\Lambda\,\approx\,-0.1173}$}}
\put(-59.5,40.0){\makebox(0,0)[l]{$^{x_0\,\approx\,-0.4566}$}}
\put(-85,-10.0){\makebox(0,0)[l]{\small (f)}}
\end{picture}
\vspace*{10pt}
\caption{Numerical solutions for $(1-x)\hat\rho(x)$ for various values of
$\tfrac{Q}{N}$. The number $L$ of `particles' for doing the numerics is always
$2000$. The solid line represents an interpolating function obtained by
numerically expanding the solutions $\hat\rho$ in terms of the Jacobi
polynomials. Notice that the analytical result for $\tfrac{Q}{N}=0$ is
$\hat\rho(x)=\frac{2}{\pi}\approx0.6366$,
$\tilde\Lambda=\frac{7}{12}-\log2\approx-0.1098$
and $x_0=-1$. The small oscillations are an artifact of the discretisation:
They
become smaller as $L$ is taken larger. These results agree with those found
previously in \cite{Berenstein:2007wz,Berenstein:2010dg}.}
\label{fig:LogZ}
\end{figure}
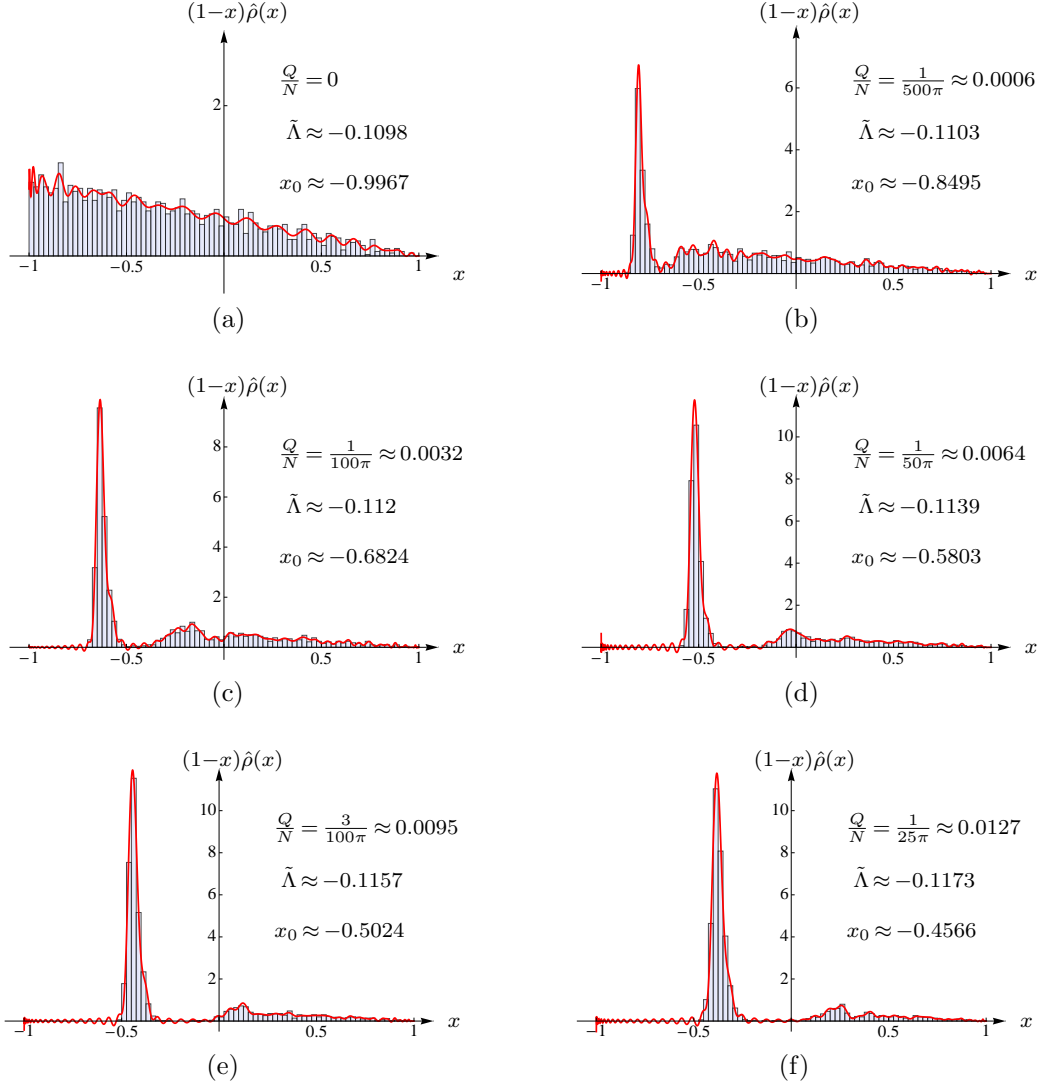

Let us first make a change of coordinates according to $x:=\cos2\theta$, so
that the normalisation and the integral equation (\ref{eq:eom-log-case-theta0})
read as ($x_0:=\cos2\theta_0$):
\begin{subequations}\label{eq:log-var-x}
\begin{eqnarray}
 0 \!&=&\! \int_{x_0}^1\!\dd x\,(1-x)\,\hat\rho(x) - \frac{4}{\pi}~,
        \label{eq:log-var-x-constraint}\\
 0 \!&=&\! \tilde\Lambda-\frac{Q}N\log(1+x)-
  \frac{\pi}{16}\int_{x_0}^1\!\dd x'\,(1-x')\,\tilde K_I(x,x')\,\hat\rho(x')~.
 \label{eq:log-var-x-rho}
 \end{eqnarray}
\end{subequations}
These equations can be obtained from variations of\footnote{Notice that
$\tilde\cH_f$ is basically a redefinition of $\cH_f$.}
\begin{equation}
 \begin{aligned}
   \tilde\cH_f \ &=\
  \tilde\Lambda\left(\int_{-1}^1\!\dd
x\,(1-x)\,\hat\rho(x)-\frac{4}{\pi}\right)
-\frac{Q}N\int_{-1}^1\!dx(1-x)\log(1+x)\hat\rho(x)\\
&\kern1.5cm -\frac{\pi}{32}\int_{-1}^1\!\dd x\,(1-x)\!\int_{-1}^1\!\dd
x'\,(1-x')\,\tilde K_I(x,x')\,\hat\rho(x)\,\hat\rho(x')~.
 \end{aligned}
\end{equation}
Note that the full range $-1\leq x\leq 1$ is used in $\tilde\cH_f$. The
restriction to $x_0\leq x\leq 1$ will be produced for configurations of
$\hat\rho$ that exactly vanish for $x<x_0$.

To extremise $\hat\cH$, we can discretise the problem by thinking of
a large number $L$ of `particles' in the interval $[-1,1]$ and trading back
integrals into sums
\begin{equation}
 \int_{-1}^1\!\dd x\,(1-x)\,\hat\rho(x)\ \to\ \frac{4}{\pi L}\sum_{i=1}^L
\end{equation}
and hence,
\begin{equation}
   \tilde\cH_f \ =\
     -\frac{Q}N\frac{4}{\pi L}\sum_{i=1}^L\log(1+x_i)
   -\frac{1}{2 \pi L^2}\sum_{i,j=1}^L\tilde K_I(x_i,x_j)~.
\end{equation}
To obtain the extrema of $\cH_f$, we now perform Monte Carlo simulations
with a modified Metropolis criterion (i.e.~we only accept new configurations if
$\delta\tilde\cH_f<0$). The big advantage of the numerical approach considered
here when compared to the one for the original Hamiltonian \eqref{eq:Hdiscrete}
is that we have effectively reduced the six-dimensional problem to a
one-dimensional problem. However, there is also a slight disadvantage in that
the interaction term appearing in $\tilde\cH_f$ is more complicated. Notice
that
the numerical value for the Lagrange multiplier $\tilde \Lambda$ (and thus
$\Lambda$ via \eqref{eq:eom-log-case-2}) can be found from the discretised
version of
\eqref{eq:log-var-x}:
\begin{equation}
 0\ =\ \tilde\Lambda-\frac{Q}N\log(1+x_i)-
\frac{1}{4L}\sum_{j=1}^L\tilde K_I(x_i,x_j)~.
\end{equation}
Notice also that this equation can be used as a criterion to estimate how
good our numerical solutions are (since $\tilde\Lambda$ should be constant for
all $x_i$). In Figure \ref{fig:LogZ}, we have depicted our numerical results
for various values of $\tfrac{Q}{N}$. The number $L$ of `particles' was chosen
to be $2000$.

\section{Conclusions and outlook}\label{sec:Conclusions}

We have considered a particular quantum mechanics model of commuting matrices
which is believed to describe 1/8 BPS states in $\sSU(N)$ $\CN=4$ SYM theory at
strong coupling. The wavefunctions of this model are expected to be dual to
type
IIB supergravity solutions that asymptotically approach AdS$_5\times S^5$. The
probability densities for the wavefunctions are interpreted as partition
functions of certain matrix models. Different wavefunctions correspond to
different potentials in the associated matrix model Hamiltonian.

Specifically, we focused on the large-$N$ or thermodynamic limit in which the
probability densities are  dominated by the saddle points of the matrix model.
Then, we have  solved analytically the saddle point equations for a family of
wavefunctions that are in correspondence with specific LLM geometries.

The starting point of our consideration was the ansatz \eqref{eq:AnsatzRho},
where it is assumed that the probability densities are supported on particular
hypersurfaces in $\IR^6$. We then constructed a perturbative approach that
allows for an analytical treatment of the saddle point equations. We used
monomial potentials to illustrate how the perturbative method works. By
perturbing around the ground state solution, we constructed the solution up to
and including second order in the deformation parameter.

We then considered a logarithmic potential, which does not admit a perturbative
solution to its saddle point equations. In spite of that, we could also obtain
partial analytical results in this case. We found an exact solution
\eqref{eq:theorad}, which holds true for any value of $Q$, for the radial
function of the ansatz \eqref{eq:AnsatzRho} and we reduced the problem to a
linear integral equation for the density $\hat \rho$
\eqref{eq:eom-log-case-theta0}. To provide a full analytical solution in the
logarithmic case, one should diagonalise the integral kernel given in
\eqref{eq:eom-log-case-2} for a generic interval $0\leq\theta\leq\theta_0$ with
$\theta_0<\tfrac{\pi}{2}$. Finally, we have compared all our analytical results
against numerical simulations and found very good agreement.

The main issue which we have not discussed here but we hope to report on in the
future concerns the extraction of geometry from our analytic solutions. Recall
that for the ground state solution, the saddle point configuration is given by
a
uniform probability distribution supported on a five-sphere in $\IR^6$ and an
explicit geometrical meaning to this five-sphere was given in
\cite{Berenstein:2005aa,Berenstein:2005jq}. The generalization of this
identification is not that obvious for the excited wavefunctions. To begin
with,
the dual LLM geometries do not have a factorised five-dimensional compact
geometry. Put it differently, the compact factor is different for different
sections of the LLM geometry. At least one would expect to be able to reproduce
the geometry of the section corresponding to the LLM plane. Then, there is also
the question of how to read geometry out of the distribution of eigenvalues.
There is obviously the hypersurface where the density is supported, but the
density function must also enter somehow. For instance, although the supporting
hypersurface in the logarithmic is a five-sphere, the compact factor in the LLM
plane of the annular LLM solution ought to be something else.

It would also be interesting to extend our perturbative method to solve
analytically the saddle point equations of more general wavefunctions. For
instance, potentials that depend also on other holomorphic coordinates are
supposed to describe wavefunctions of 1/4 and 1/8 BPS states. This would shed
light on the AdS/CFT dictionary for those cases (see e.g.~\cite{Chen:2007du}
and references therein). In addition, it would be very interesting to use our
perturbative method to characterise excited wavefunctions in other matrix
quantum mechanical systems
\cite{Berenstein:2005ek,Berenstein:2006yy,Berenstein:2007wi,Berenstein:2007kq},
which according to the AdS/CFT correspondence are dual to supergravity
solutions with other asymptotics (e.g.~AdS$_5\times T^{1,1}$).

\bigskip
\bigskip
\noindent
{\bf Acknowledgements.}
We are very grateful to D.~Berenstein for important discussions and for sharing
the {\ttfamily C++} code used to create Figure \ref{fig:logx2}a). D.H.C.
was supported in parts by the Seventh Framework Programme under grant agreement
number PIEF--GA--2008--220702. M.W. was
supported by an STFC Postdoctoral Fellowship and by a Senior Research
Fellowship
at the Wolfson College, Cambridge, U.K.

\addtocontents{toc}{\hspace{-0.61cm}Appendices}
\appendix
\bigbreak\bigskip\centerline{{\bf Appendices}}\nobreak
\vspace*{10pt}

\renewcommand{\thesection}{\Alph{section}.}
\setcounter{subsection}{0} \setcounter{equation}{0}
\renewcommand{\thesubsection}{\Alph{subsection}}
\renewcommand{\theequation}{\thesubsection.\arabic{equation}}

\subsection{Eigenvalues and eigenfunctions}\label{app:AppendixKernel}

In this appendix, we shall derive the eigenvalues and eigenfunctions of the
kernels $K_{\rm I}$ and $K_{\rm II}$ introduced in \eqref{eq:Kernels}. Before
tackling the problem, we begin with some preliminary considerations.

\subsubsection{Preliminaries}

In the definition \eqref{eq:Kernels} of the kernels $K_{\rm I}$ and $K_{\rm
II}$, we are integrating over $(\alpha,\xi_1,\xi_2)$ and
$(\alpha',\xi'_1,\xi'_2)$, i.e.~over all possible positions of two points on a
three-sphere. The dependence of $\varphi$ on them is through the relative angle
between two points on the three-sphere, so we can fix one of the points
arbitrarily (say $\alpha'=0$ and $\xi'_1 =0$) and multiply by the volume,
$2\pi^2$, of the three-sphere. In the mentioned choice, also the integral over
$\xi_2$ becomes trivial and we obtain
\begin{subequations}
\begin{equation}\label{eq:Kernels-V2}
 \begin{aligned}
  K_{\rm I}(\omega,\omega') \ &=\
 \frac{2}{\pi^2}\int_0^{\pi/2}\!\!\!\!\dd\alpha\int_0^{2\pi}\!\!\!\!\dd\xi_1
 \sin\alpha\cos\alpha\, \log(1-\cos\varphi_0)~,\\
K_{\rm II}(\omega,\omega') \ &=\
\frac{2}{\pi^2}\int_0^{\pi/2}\!\!\!\!\dd\alpha\int_0^{2\pi}\!\!\!\!\dd\xi_1
 \sin\alpha\cos\alpha\,\frac{1}{1-\cos\varphi_0}
 \end{aligned}
\end{equation}
where
\begin{equation}
 \cos\varphi_0\ :=\ \cos\theta\cos\theta'\cos(\phi-\phi')+
\sin\theta\sin\theta'
\cos\alpha\cos\xi_1~.
\end{equation}
\end{subequations}

There are two (equivalent) ways of computing \eqref{eq:Kernels-V2}. Firstly, we
may expand the expressions $\log(1-\cos\varphi_0)$ and $1/(1-\cos\varphi_0)$ in
powers of $\cos\varphi_0$:
\begin{subequations}\label{eq:KernelSeries}
\begin{equation}\label{eq:KernelExpansions}
 K_{\rm I}(\omega,\omega')\ =\
 -\sum_{j=1}^\infty\frac{1}{j}\,K_j(\omega,\omega')\eand
 K_{\rm II}(\omega,\omega')\ =\ \sum_{j=0}^\infty K_j(\omega,\omega')~,
\end{equation}
with
\begin{equation}
 K_j(\omega,\omega')\ :=\ \frac{2}{\pi^2}
 \int_0^{\pi/2}\!\!\!\!\dd\alpha\int_0^{2\pi}\!\!\!\!\dd\xi_1
 \sin\alpha\cos\alpha\cos^j\varphi_0~.
\end{equation}
\end{subequations}
A short calculation reveals that
\begin{equation}\label{eq:SubKernels}
 K_j(\omega,\omega')\ =\ \frac{2}{\pi}\,
[\cos\theta\cos\theta'\cos(\phi-\phi')]^j\,
{}_2F_1\left(\frac{1-j}{2},\frac{-j}{2};2;\frac{\tan^2\theta\tan^2\theta'}{
\cos^2(\phi-\phi')}\right),
\end{equation}
where ${}_2F_1$ is the hypergeometric function. Notice that either the first or
the second argument of the hypergeometric function appearing in
\eqref{eq:SubKernels} is a non-positive integer. This implies that
${}_2F_1\big(\frac{1-j}{2},\frac{-j}{2};2;\cdot\big)$ is actually
a polynomial in the last argument.

As an alternative to the above expansions, one may perform the integrals in
\eqref{eq:Kernels-V2} directly. In particular, they can be re-written as
\begin{subequations}
\begin{equation}\label{eq:Kernels-V3}
\begin{aligned}
 K_{\rm I}(\omega,\omega') \ &=\
  \frac{2}{\pi^2}\int_0^1\!\!\!\dd t\int_0^{2\pi}\!\!\!\dd\xi_1\,
  t \log(A-B\, t\cos\xi_1)~,\\
 K_{\rm II}(\omega,\omega') \ &=\
  \frac{2}{\pi^2}\int_0^1\!\!\!\dd t\int_0^{2\pi}\!\!\!\dd\xi_1
  \,\frac{t}{A-B\, t\cos\xi_1}~,
\end{aligned}
\end{equation}
where
\begin{equation}\label{eq:DefinitionAB}
 A\ =\ A(\omega,\omega')\ :=\ 1-\cos\theta\cos\theta'\cos(\phi-\phi')
 \eand
 B\ =\ B(\omega,\omega')\ :=\ \sin\theta\sin\theta'~.
\end{equation}
\end{subequations}
Notice that $|B/A|\leq 1$ for all $(\omega,\omega')$. Using ($|\alpha|\leq1$)
\begin{equation}
\begin{aligned}
 \int_0^{2\pi}\!\!\!\!\dd x\,\log(1-\alpha\cos x)\ &=\
 2\pi\log\tfrac12\big(1+\sqrt{1-\alpha^2}\big)~,\\
 \int_0^{2\pi}\!\!\!\!\dd x\,\frac{1}{1-\alpha\cos x}\ &=\
 \frac{2\pi}{\sqrt{1-\alpha^2}}~,
\end{aligned}
\end{equation}
the kernels \eqref{eq:Kernels-V3} are given by
\begin{equation}\label{eq:Kernels-V4}
\begin{aligned}
 K_{\rm I}(\omega,\omega') \ &=\ \frac{2}{\pi}\left[\log
  \frac12\left(A+\sqrt{A^2-B^2}\right)
  -\frac12+\frac{A}{B^2}\left(A-\sqrt{A^2-B^2}\right)\right]\\
  &=\ \frac{2}{\pi}\left[\log A-
  \frac{B^2}{8A^2}\,{}_3F_2\left(1,1,\frac32;2,3;\frac{B^2}{A^2}\right)\right],
   \\
  K_{\rm II}(\omega,\omega') \ &=\
  \frac{2}{\pi}\,\frac{2}{B^2}\left(A-\sqrt{A^2-B^2}\right)\\
  &=\
  \frac{2}{\pi}\,\frac{1}{A}\,{}_2F_1\left(\frac12,1;2;\frac{B^2}{A^2}\right),
 \end{aligned}
\end{equation}
where ${}_3F_2$ is a generalised hypergeometric function.

Notice that as a by-product we have obtained the non-trivial identities
\begin{equation}
\begin{aligned}
 -\sum_{j=1}^\infty\frac1j\,x^j
 {}_2F_1\left(\frac{1-j}{2},\frac{-j}{2};2;y^2\right)\ &=\
 \log(1-x)-
 \frac{x^2y^2}{8(1-x)^2}\,{}_3F_2\left(1,1,\frac32;2,3;\frac{x^2y^2}{(1-x)^2}
 \right),\\
 \sum_{j=0}^\infty x^j
 {}_2F_1\left(\frac{1-j}{2},\frac{-j}{2};2;y^2\right)\ &=\
 \frac{1}{1-x}\,{}_2F_1\left(\frac12,1;2;\frac{x^2y^2}{(1-x)^2}\right)
 \end{aligned}
\end{equation}
between hypergeometric functions.

\subsubsection{Eigenvalue problem}

Let us now derive the eigenvalues and eigenfunctions for the kernels
$K_{\rm I}$ and $K_{\rm II}$. To this end, the form \eqref{eq:KernelSeries}
turns out to be very adequate, since the sub-kernels \eqref{eq:SubKernels} can
actually be diagonalised simultaneously.\footnote{Therefore, any kernel of
the form $\sum_j a_j K_j$ with some constant coefficients $a_j$ can be
diagonalised in this manner.} In fact,
\begin{subequations}\label{eq:SpecProbSubKern}
\begin{equation}
 \int\!\dd\Omega'_2\,
 K_j(\omega,\omega') \Psi_{m,a}(\omega')\ =\
 \underbrace{\frac{(1+(-1)^{|m|+j})\,2^{-j}\,
 \Gamma(j+1)}{\Gamma\big(\tfrac{j}{2}-\tfrac{|m|}{2}-a+1\big)
 \Gamma\big(\tfrac j2 +\tfrac{|m|}{2}+a+3\big)}}_{=:\ \lambda^{m,a}_j}\,
\Psi_{m,a}(\omega)~,
\end{equation}
where
\begin{equation}
 \Psi_{m,a}(\omega)\ =\ \cos^{|m|}\theta\,
P^{(1,|m|)}_a(\cos2\theta)\,\exp(\di m\phi)~
\end{equation}
\end{subequations}
for $m\in\IZ$ and $a\in\IN_0$. Here, $P_a^{(\alpha,\beta)}$ are the Jacobi
polynomials (see Appendix \ref{app:Jacobi}) and $\Gamma$ denotes the Gamma
function.

These eigenfunctions form a complete orthogonal basis for functions defined on
the hemisphere given by $\omega$; see Figure \ref{fig:FiveSphere}. Indeed,
upon introducing the delta function
\begin{equation}
 \delta(\omega,\omega')\ :=\ \frac{\pi}{2}\,\frac{1}{\sin^3\theta\cos\theta}\,
 \delta(\theta-\theta')\delta(\phi-\phi')~,\ewith
 \frac{2}{\pi}\int\!\dd\Omega'_2\,\delta(\omega,\omega')\ =\ 1~,
\end{equation}
the completeness relation is given by
\begin{equation}
 \sum_{m,a}\frac{(|m|+2a+2)(|m|+a+1)}{2a+2}\,
\Psi_{m,a}(\omega)[\Psi_{m,a}(\omega')]^*\ =\
\delta(\omega,\omega')~,
\end{equation}
where `$*$' denotes complex conjugation. This follows from the completeness
relation \eqref{eq:CompletenessJP} of the Jacobi polynomials. Furthermore, the
orthogonality relation reads as
\begin{equation}
 \frac{2}{\pi}\int\!
 \dd\Omega_2\,\Psi_{m,a}(\omega)[\Psi_{n,b}(\omega)]^*\ =\
 \frac{2a+2}{(|m|+2a+2)(|m|+a+1)}\,\delta_{mn}\delta_{ab}~,
\end{equation}
where $\delta$ is the Kronecker delta. This expression can be obtained from the
orthogonality relation \eqref{eq:OrthogonalityJP} of the Jacobi polynomials.
Altogether, we may obtain an orthonormal basis by
setting
\begin{equation}
 \Psi_{m,a}(\omega)\ \mapsto\ \Psi^0_{m,a}(\omega)\ :=\
  \sqrt{\frac{(|m|+2a+2)(|m|+a+1)}{2a+2}}\,\Psi_{m,a}(\omega)~
\end{equation}
for which we have
\begin{equation}
  \sum_{m,a}\Psi^0_{m,a}(\omega)[\Psi^0_{m,a}(\omega')]^*\ =\
\delta(\omega,\omega')\eand
 \frac{2}{\pi}\int\!
 \dd\Omega_2\,\Psi^0_{m,a}(\omega)[\Psi^0_{n,b}(\omega)]^*\ =\
 \delta_{mn}\delta_{ab}~.
\end{equation}

Next we would like to derive the eigenvalues  $\lambda_{\rm I,II}^{m,a}$. They
are given by summing up the eigenvalues $\lambda_j^{m,a}$ appearing in
\eqref{eq:SpecProbSubKern}:
\begin{equation}\label{eq:EigenvaluesFromSubKernels}
\begin{aligned}
 \lambda_{\rm I}^{m,a}\ &=\ -\sum_{j=1}^\infty\frac1j\,\lambda_j^{m,a}\ =\
  \begin{cases}
   {\displaystyle\frac{7}{12}} - \log 2 &\efor (m,a)\ =\ (0,0)~,\\
   -24\,{\displaystyle\frac{(|m|+2a-1)!}{(|m|+2a+4)!}} &\efor (m,a)\ \neq\
    (0,0)~,\end{cases}\\
 \lambda_{\rm II}^{m,a}\ &=\ \sum_{j=0}^\infty\lambda_j^{m,a}\ =\
 8\,\frac{(|m|+2a)!}{(|m|+2a+3)!}~.
\end{aligned}
\end{equation}
Using the eigenvalues $\lambda_{\rm I,II}^{m,a}$ and the normalised
eigenfunctions $\Psi_{m,a}^0$, the integral kernels $K_{\rm I}$ and $K_{\rm
II}$
may be expressed as
\begin{equation}\label{eq:KernelEFExpansion}
 K_{\rm I,II}(\omega,\omega')\ =\ \frac{2}{\pi}\sum_{m,a}\lambda_{\rm
I,II}^{m,a}\, \Psi^0_{m,a}(\omega)[\Psi^0_{m,a}(\omega')]^*~.
\end{equation}

Notice that upon using \eqref{eq:Kernels-V4}, one easily computes the traces
$\mbox{tr}\,K_{\rm I}$ and $\mbox{tr}\,K_{\rm II}$. They can also be
obtained by summing up the eigenvalues
\eqref{eq:EigenvaluesFromSubKernels}:
\begin{equation}
 \mbox{tr}\,K_{\rm I,II}\ =\ \int\!
 \dd\Omega_2\, K_{\rm I,II}(\omega,\omega)\ =\
 \sum_{m,a}\lambda_{\rm I,II}^{m,a} \ =\
  \begin{cases} -\log2~,\\ 4~. \end{cases}
\end{equation}

\subsection{Jacobi polynomials}\label{app:Jacobi}

The Jacobi polynomials $P_a^{(\alpha,\beta)}$ for $\alpha,\beta\in\IR$ (with
$\alpha,\,\beta>-1$) and
$a\in\IN_0$ are solutions to the ordinary differential equation
\begin{equation}
 (1-x^2)y''+(\beta-\alpha-(\alpha+\beta+2)x)y'+a(a+\alpha+\beta+1)y\ =\ 0~
\end{equation}
and they can be obtained from the hypergeometric function ${}_2F_1$ according
to
\begin{equation}
   P_a^{(\alpha,\beta)}(x)\ =\
   \frac{(\alpha+1)_a}{a!}\,
{}_2F_1\left(-a,1+\alpha+\beta+a;\alpha+1;\frac{1-x}{2}\right),
\end{equation}
where $(a)_n:=\Gamma(a+n)/\Gamma(a)$ is the Pochhammer symbol. We are
particularly interested in the polynomials $P^{(1,m)}_a$ which are:
\begin{equation}
 \begin{aligned}
   P^{(1,m)}_0(x)\ &=\ 1~,\\
   P^{(1,m)}_1(x)\ &=\ 2+ \tfrac12 (3 + m) (x - 1)~,\\
   P^{(1,m)}_2(x)\ &=\ 3 +\tfrac32 (4 + m) (x-1) + \tfrac18 (4 + m) (5 + m)
                    (x-1)^2~,\\
   P^{(1,m)}_3(x)\ &=\ 4 + 3 (5 + m) (x-1) + \tfrac12 (5 + m) (6 + m)
(x-1)^2\\
 &~~~~~~~~~+\,\tfrac{1}{48} (5 + m) (6 + m) (7 + m) (x-1)^3~,\\
    &~\,\vdots
 \end{aligned}
\end{equation}

Furthermore, the Jacobi polynomials form an orthogonal basis with
\begin{equation}\label{eq:OrthogonalityJP}
\int_{-1}^1\!\!\dd x\,(1-x)^\alpha(1+x)^\beta
 P_a^{(\alpha,\beta)}(x) P_b^{(\alpha,\beta)}(x)\ =\
\frac{2^{\alpha+\beta+1}}{2a+\alpha+\beta+1}\frac{(a+1)_\alpha}{
(\beta+a+1)_\alpha}\,\delta_{ab}~.
\end{equation}
The completeness relation is given by
\begin{equation}\label{eq:CompletenessJP}
 \sum_{a=0}^\infty
\frac{(\alpha+\beta+2a+1)(\beta+a+1)_\alpha}{
(a+1)_\alpha}\,
 P_a^{(\alpha,\beta)}(x)P_a^{(\alpha,\beta)}(y)\ =\
 \frac{2^{\alpha+\beta+1}}{(1-x)^\alpha(1+x)^\beta}\,\delta(x-y)~
\end{equation}
for $|x|<1$ and $|y|<1$.

Finally, we record the following useful relations:
\begin{equation}
 P^{(\alpha,\beta)}_a(-x)\ =\ (-1)^a P_a^{(\beta,\alpha)}(x)
\end{equation}
and
\begin{equation}
 \frac{\dd^k}{\dd x^k}P^{(\alpha,\beta)}_a(x)\ =\
\frac{(\alpha+\beta+a+1)_k}{2^k}\,
P^{(\alpha+k,\beta+k)}_{a-k}(x)~.
\end{equation}


\bigskip

\begin{thebibliography}{10}
\ifx\href\asklfhas\newcommand{\href}[2]{#2}\fi
\ifx\arxivref\asklfhas\newcommand{\arxivref}[2]{\href{http://arxiv.org/abs/#1}%
{#2}}\fi
\ifx\doiref\asklfhas\newcommand{\doiref}[2]{\href{http://dx.doi.org/#1}{#2}}\fi
\raggedright
\small
\parskip 0pt

%%CITATION = HEP-TH/9711200;%%
\bibitem{Maldacena:1997re}
J.~M.~Maldacena,
\textit{``{The large-N limit of superconformal field theories and
  supergravity}''},
\textsf{\doiref{10.1023/A:1026654312961}{Adv.~Theor.~Math.~Phys.~2,~231~(1998)%
}},
\texttt{\arxivref{hep-th/9711200}{hep-th/9711200}}.

%%CITATION = HEP-TH/9802109;%%
\bibitem{Gubser:1998bc}
S.~S.~Gubser, I.~R.~Klebanov and A.~M.~Polyakov,
\textit{``{Gauge theory correlators from non-critical string theory}''},
\textsf{\doiref{10.1016/S0370-2693(98)00377-3}{Phys.~Lett.~B428,~105~(1998)}},
\texttt{\arxivref{hep-th/9802109}{hep-th/9802109}}.

%%CITATION = HEP-TH/9802150;%%
\bibitem{Witten:1998qj}
E.~Witten,
\textit{``{Anti-de Sitter space and holography}''},
\textsf{Adv.~Theor.~Math.~Phys.~2,~253~(1998)},
\texttt{\arxivref{hep-th/9802150}{hep-th/9802150}}.

%%CITATION = HEP-TH/0507203;%%
\bibitem{Berenstein:2005aa}
D.~Berenstein,
\textit{``{Large-N BPS states and emergent quantum gravity}''},
\textsf{\doiref{10.1088/1126-6708/2006/01/125}{JHEP~0601,~125~(2006)}},
\texttt{\arxivref{hep-th/0507203}{hep-th/0507203}}.

%%CITATION = 0805.4658;%%
\bibitem{Berenstein:2008eg}
D.~E.~Berenstein, M.~Hanada and S.~A.~Hartnoll,
\textit{``{Multi-matrix models and emergent geometry}''},
\textsf{\doiref{10.1088/1126-6708/2009/02/010}{JHEP~0902,~010~(2009)}},
\texttt{\arxivref{0805.4658}{arxiv:0805.4658}}.

%%CITATION = HEP-TH/0511104;%%
\bibitem{Berenstein:2005ek}
D.~Berenstein and D.~H.~Correa,
\textit{``{Emergent geometry from $q$-deformations of $\CN=4$ super
  Yang--Mills}''},
\textsf{\doiref{10.1088/1126-6708/2006/08/006}{JHEP~0608,~006~(2006)}},
\texttt{\arxivref{hep-th/0511104}{hep-th/0511104}}.

%%CITATION = HEP-TH/0605220;%%
\bibitem{Berenstein:2006yy}
D.~Berenstein and R.~Cotta,
\textit{``{Aspects of emergent geometry in the AdS/CFT context}''},
\textsf{\doiref{10.1103/PhysRevD.74.026006}{Phys.~Rev.~D74,~026006~(2006)}},
\texttt{\arxivref{hep-th/0605220}{hep-th/0605220}}.

%%CITATION = 0710.2086;%%
\bibitem{Berenstein:2007wi}
D.~Berenstein,
\textit{``{Strings on conifolds from strong coupling dynamics, part I}''},
\textsf{\doiref{10.1088/1126-6708/2008/04/002}{JHEP~0804,~002~(2008)}},
\texttt{\arxivref{0710.2086}{arxiv:0710.2086}}.

%%CITATION = 0711.3026;%%
\bibitem{Berenstein:2007kq}
D.~E.~Berenstein and S.~A.~Hartnoll,
\textit{``{Strings on conifolds from strong coupling dynamics: quantitative
  results}''},
\textsf{\doiref{10.1088/1126-6708/2008/03/072}{JHEP~0803,~072~(2008)}},
\texttt{\arxivref{0711.3026}{arxiv:0711.3026}}.

%%CITATION = HEP-TH/0702090;%%
\bibitem{Berenstein:2007wz}
D.~Berenstein and R.~Cotta,
\textit{``{A Monte Carlo study of the AdS/CFT correspondence: An exploration of
  quantum gravity effects}''},
\textsf{\doiref{10.1088/1126-6708/2007/04/071}{JHEP~0704,~071~(2007)}},
\texttt{\arxivref{hep-th/0702090}{hep-th/0702090}}.

%%CITATION = HEP-TH/0509015;%%
\bibitem{Berenstein:2005jq}
D.~Berenstein, D.~H.~Correa and S.~E.~Vazquez,
\textit{``{All loop BMN state energies from matrices}''},
\textsf{\doiref{10.1088/1126-6708/2006/02/048}{JHEP~0602,~048~(2006)}},
\texttt{\arxivref{hep-th/0509015}{hep-th/0509015}}.

%%CITATION = HEP-TH/0111222;%%
\bibitem{Corley:2001zk}
S.~Corley, A.~Jevicki and S.~Ramgoolam,
\textit{``{Exact correlators of giant gravitons from dual $\CN = 4$ SYM
  theory}''},
\textsf{Adv.~Theor.~Math.~Phys.~5,~809~(2002)},
\texttt{\arxivref{hep-th/0111222}{hep-th/0111222}}.

%%CITATION = HEP-TH/0403110;%%
\bibitem{Berenstein:2004kk}
D.~Berenstein,
\textit{``{A toy model for the AdS/CFT correspondence}''},
\textsf{\doiref{10.1088/1126-6708/2004/07/018}{JHEP~0407,~018~(2004)}},
\texttt{\arxivref{hep-th/0403110}{hep-th/0403110}}.

%%CITATION = HEP-TH/0507070;%%
\bibitem{Takayama:2005yq}
Y.~Takayama and A.~Tsuchiya,
\textit{``{Complex matrix model and fermion phase space for bubbling AdS
  geometries}''},
\textsf{\doiref{10.1088/1126-6708/2005/10/004}{JHEP~0510,~004~(2005)}},
\texttt{\arxivref{hep-th/0507070}{hep-th/0507070}}.

%%CITATION = HEP-TH/0507124;%%
\bibitem{Donos:2005vm}
A.~Donos, A.~Jevicki and J.~P.~Rodrigues,
\textit{``{Matrix model maps in AdS/CFT}''},
\textsf{\doiref{10.1103/PhysRevD.72.125009}{Phys.~Rev.~D72,~125009~(2005)}},
\texttt{\arxivref{hep-th/0507124}{hep-th/0507124}}.

%%CITATION = 0911.4817;%%
\bibitem{Koch:2009gq}
R.~d.~M.~Koch and J.~Murugan,
\textit{``{Emergent space-time}''},
\texttt{\arxivref{0911.4817}{arxiv:0911.4817}}.

%%CITATION = HEP-TH/0409174;%%
\bibitem{Lin:2004nb}
H.~Lin, O.~Lunin and J.~M.~Maldacena,
\textit{``{Bubbling AdS space and 1/2 BPS geometries}''},
\textsf{\doiref{10.1088/1126-6708/2004/10/025}{JHEP~0410,~025~(2004)}},
\texttt{\arxivref{hep-th/0409174}{hep-th/0409174}}.

%%CITATION = HEP-TH/0612014;%%
\bibitem{Vazquez:2006id}
S.~E.~Vazquez,
\textit{``{Reconstructing 1/2 BPS space-time metrics from matrix models and
  spin chains}''},
\textsf{\doiref{10.1103/PhysRevD.75.125012}{Phys.~Rev.~D75,~125012~(2007)}},
\texttt{\arxivref{hep-th/0612014}{hep-th/0612014}}.

%%CITATION = HEP-TH/0703068;%%
\bibitem{Chen:2007gh}
H.-Y.~Chen, D.~H.~Correa and G.~A.~Silva,
\textit{``{Geometry and topology of bubble solutions from gauge theory}''},
\textsf{\doiref{10.1103/PhysRevD.76.026003}{Phys.~Rev.~D76,~026003~(2007)}},
\texttt{\arxivref{hep-th/0703068}{hep-th/0703068}}.

%%CITATION = 0806.0685;%%
\bibitem{Koch:2008ah}
R.~de~Mello~Koch,
\textit{``{Geometries from Young diagrams}''},
\textsf{\doiref{10.1088/1126-6708/2008/11/061}{JHEP~0811,~061~(2008)}},
\texttt{\arxivref{0806.0685}{arxiv:0806.0685}}.

%%CITATION = 0911.0967;%%
\bibitem{deMelloKoch:2009zm}
R.~de~Mello~Koch, T.~K.~Dey, N.~Ives and M.~Stephanou,
\textit{``{Hints of Integrability Beyond the Planar Limit}''},
\textsf{\doiref{10.1007/JHEP01(2010)014}{JHEP~1001,~014~(2010)}},
\texttt{\arxivref{0911.0967}{arxiv:0911.0967}}.

%%CITATION = 1003.4190;%%
\bibitem{Lin:2010sb}
H.~Lin, A.~Morisse and J.~P.~Shock,
\textit{``{Strings on Bubbling Geometries}''},
\textsf{\doiref{10.1007/JHEP06(2010)055}{JHEP~1006,~055~(2010)}},
\texttt{\arxivref{1003.4190}{arxiv:1003.4190}}.

%%CITATION = 0801.2739;%%
\bibitem{Berenstein:2008jn}
D.~Berenstein, R.~Cotta and R.~Leonardi,
\textit{``{Numerical tests of AdS/CFT at strong coupling}''},
\textsf{\doiref{10.1103/PhysRevD.78.025008}{Phys.~Rev.~D78,~025008~(2008)}},
\texttt{\arxivref{0801.2739}{arxiv:0801.2739}}.

%%CITATION = 1001.4509;%%
\bibitem{Berenstein:2010dg}
D.~Berenstein and Y.~Nakada,
\textit{``{The shape of emergent quantum geometry from an $\CN=4$ SYM
  minisuperspace approximation}''},
\texttt{\arxivref{1001.4509}{arxiv:1001.4509}}.

%%CITATION = CMPHA,59,35;%%
\bibitem{Brezin:1977sv}
E.~Brezin, C.~Itzykson, G.~Parisi and J.~B.~Zuber,
\textit{``{Planar diagrams}''},
\textsf{\doiref{10.1007/BF01614153}{Commun.~Math.~Phys.~59,~35~(1978)}}.

\bibitem{Tricomi:1957}
F.~G.~Tricomi,
\textit{``Integral equations''},
Pure Appl.\ Math.\ V, Interscience, London, 1957.

%%CITATION = 0704.2233;%%
\bibitem{Chen:2007du}
B.~Chen, S.~Cremonini, A.~Donos, F.-L.~Lin, H.~Lin, J.~T.~Liu, D.~Vaman and
  W.-Y.~Wen,
\textit{``{Bubbling AdS and droplet descriptions of BPS geometries in IIB
  supergravity}''},
\textsf{\doiref{10.1088/1126-6708/2007/10/003}{JHEP~0710,~003~(2007)}},
\texttt{\arxivref{0704.2233}{arxiv:0704.2233}}.

\end{thebibliography}
\end{document}